\documentstyle[12pt,epsf,amsmath,amsthm,amssymb]{article}

\let\LARGE=\Large
\let\Large=\large
\let\large=\normalsize
\makeatletter
\def\verbatim{\interlinepenalty\@M \@verbatim
  \leftskip\@totalleftmargin\advance\leftskip2pc
  \frenchspacing\@vobeyspaces \@xverbatim}
\makeatother
\newtheorem{defn}{Definition}
\newtheorem{thm}{Theorem}
\newtheorem{cor}{Corollary}

\def\1I{\relax{\rm 1\kern-.25em \rm l}} 

\newcommand{\unity}{\1I}
\newcommand{\ins}{{\raisebox{0.2ex}{$\,\lrcorner\:$}}}

\newcommand{\MyIm}{\Im{\mathfrak{m}}}

\newcommand{\be}[3]{\begin{equation}  \label{#1#2#3}}     
\newcommand{\ee}{ \end{equation}}
\newcommand{\ba}{\begin{array}}
\newcommand{\ea}{\end{array}}

\renewcommand{\arraystretch}{1.8}

\def\beq{\begin{equation}}
\def\eeq{\end{equation}}
\def\beqa{\begin{eqnarray}}
\def\eeqa{\end{eqnarray}}

\def\fund{  \> {\vcenter  {\vbox
              {\hrule height.6pt
               \hbox {\vrule width.6pt  height5pt
                      \kern5pt
                      \vrule width.6pt  height5pt }
               \hrule height.6pt}
                         }
                   }
           \>\> }
\def\antifund{  \> \overline{ {\vcenter  {\vbox
              {\hrule height.6pt
               \hbox {\vrule width.6pt  height5pt
                      \kern5pt
                      \vrule width.6pt  height5pt }
               \hrule height.6pt}
                         }
                   } }
           \>\> }

\def\sym{  \> {\vcenter  {\vbox
              {\hrule height.6pt
               \hbox {\vrule width.6pt  height5pt
                      \kern5pt
                      \vrule width.6pt  height5pt
                      \kern5pt
                      \vrule width.6pt height5pt}
               \hrule height.6pt}
                         }
              }
           \>\> }

\def\symbar{  \> \overline{ {\vcenter  {\vbox
              {\hrule height.6pt
               \hbox {\vrule width.6pt  height5pt
                      \kern5pt
                      \vrule width.6pt  height5pt
                      \kern5pt
                      \vrule width.6pt height5pt}
               \hrule height.6pt}
                         }
              }
           } \>\> }
\def\anti{ \> {\vcenter  {\vbox
              {\hrule height.6pt
               \hbox {\vrule width.6pt  height5pt
                      \kern5pt
                      \vrule width.6pt  height5pt }
               \hrule height.6pt
               \hbox {\vrule width.6pt  height5pt
                      \kern5pt
                      \vrule width.6pt  height5pt }
               \hrule height.6pt}
                         }
              }
           \>\> }

\def\href#1#2{#2}

\begin{document}

\thispagestyle{empty}
\rightline{HUB-EP-98/70}
\rightline{hep-th/9810254}
\vspace{2truecm}
\centerline{\bf \LARGE  $N=1$ Supersymmetric Gauge Theories and Supersymmetric
3-cycles}

\vspace{1.5truecm}
\centerline{\bf 
Andreas Karch\footnote{karch@physik.hu-berlin.de}\ , \ 
Dieter L\"ust\footnote{luest@physik.hu-berlin.de}\ and \
Andr\'e Miemiec\footnote{miemiec@physik.hu-berlin.de}}

\vspace{.5truecm}
{\em 
\centerline{Humboldt-Universit\"at, Institut f\"ur Physik,
D-10115 Berlin, Germany}}

\vspace{1.0truecm}
\begin{abstract}
In this paper we discuss the strong coupling limit
 of chiral $N=1$ supersymmetric
gauge theory via their embedding into M-theory. In particular we focus
on the brane box models of Hanany and Zaffaroni and show that after
a T-duality transformation their M-theory embedding is described
by supersymmetric 3-cycles; its geometry will encode
the holomorphic non-perturbative information about the gauge theory. 
\end{abstract}

\bigskip \bigskip
\newpage

\tableofcontents\newpage


\section{Introduction}

Recently it was
demonstrated that many 
interesting non-perturbative results in superstring theory and
supersymmetric field theories can be derived from 11-dimensional M-theory.
In particular Witten \cite{witten} has shown that $N=2$ supersymmetric
gauge theories can be solved via M-theory by lifting the corresponding
10-dimensional type II brane configurations 
\cite{hanwit} to 11 dimensions. As a result the
intersection of $n$ parallel NS 5-branes and
$k$ suspended D 4-branes  is described in 11 dimensions by an 
one-dimensional complex curve, which can be seen as a two-dimensional
supersymmetric cycle embedded in the flat four-dimensional space 
${\Bbb{R}}^3\times S^1$.
This curve precisely agrees with the famous Seiberg-Witten curves \cite{sw}
of $N=2$ supersymmetric field theories.

Non-chiral $N=1$ gauge theories can be obtained by rotating one or several
of the NS 5-branes such that they intersect by a certain angle.
The corresponding continuous parameter can be regarded in field theory as
a mass parameter which explicitly breaks $N=2$ supersymmetry down to
$N=1$.
The M-theory embedding of the non-chiral $N=1$ models, constructed in this
way, leads again to supersymmetric 2-cycles, now embedded in the flat
six-dimensional space ${\Bbb{R}}^5\times S^1$ \cite{hori,witten1,9706127}. Analyzing
these curves, the form of the corresponding $N=1$ superpotentials can
be derived.

A generic way to construct {\it chiral} $N=1$ gauge theories in four dimensions
is provided by the
brane box models of Hanany and Zaffaroni \cite{hanzaf}\footnote{An alternative,
but more restricted construction of chiral $N=1$ models via
orientifolds was introduced in \cite{Landsteiner,alter}.}.
Here one deals with a type IIB configuration of intersecting NS and D 5-branes.
We will show that upon a T-duality transformation to the IIA superstring,
the M-theory lifting of the chiral $N=1$ brane box configurations leads
to supersymmetric 3-cycles suitably embedded into ${\Bbb{C}}^3$.
These supersymmetric 3-cycles precisely correspond to the $SU(3)$
special Lagrangian calibrations (SLAG), discussed e.g. in
\cite{gibpap,9803216}, which reduce the original supersymmetry by a factor 
1/8.
This is just the right amount of supersymmetry breaking for a generic
chiral $N=1$ gauge theory which cannot be smoothly deformed into 
a $N=2$ gauge theory.

Following \cite{witten} one can study quantum effects in the corresponding
gauge theory by analyzing how the branes bend each other. Especially it
is easy to study the $\beta$ function: the gauge coupling is encoded
in the distance between two NS branes or in the area of the box in the
case of boxes respectively. So the shape of the bent branes directly gives
some information about the running gauge coupling. In order to obtain
the exact quantum information (or at least the information protected by
holomorphy) one can lift the brane configuration to M-theory and use
11d SUGRA to solve it. For IIA brane configurations like those studied
in \cite{witten} this is straight forward, since 11d SUGRA on a circle
is dual to IIA.

The boxes of \cite{hanzaf} live in IIB theory, so in order to perform the
M-theory lift on has to use the relation that IIB on a circle is M-theory
on a torus, so we have to compactify one of the common wordvolume
directions. Like in the 5d case studied in \cite{kol} this means
that we are really solving the 4d theory on $R^3 \times S^1$.
In the limit where the area of the M-theory torus shrinks to zero
or grows to infinity we recover the $N=1$ $d=4$ and $N=2$ $d=3$ gauge
theories respectively.
We will show that this way all the holomorphic information about these
gauge theories is encoded in the geometry of a SUSY 3-cycle. Especially
we expect the superpotential to correspond to the volume and the
gauge couplings on the Coulomb branches to be related to periods of
the cycle. The possible 3-cycles for given boundary values will encode
the vacuum structure of the theory. Gauge theories which dynamical
SUSY breaking will correspond to situation, where the minimal area
cycle for the given boundary problem is not a SUSY 3-cycle.

In Section 2 we will review Witten's rediscovery of
the Seiberg-Witten curve in terms of the lift to M-theory. In Section 3
we will introduce the classical brane box setups and show how they
relate to 3-cycles. We will develop some tools which we believe are
very helpful in constructing the 3-cycles. For the special
case of finite theories and theories that satisfy the uniform
bending requirement of \cite{gimongremm} we are able to perform the lift
explicitly. However the corresponding cycles turn out to be superpositions
of special 3-cycles that are of the form 2-cycle times line.
We end with some preliminary results about more general cases.


\section{{\it\bf N=2} Gauge Theories and 2-cycles}

In order to understand better the supersymmetric
3-cycles in  the $N=1$ brane box models,
let us first recall the M-theory embedding of the $N=2$ models which
leads to supersymmetric 2-cycles.
The four-dimensional $N=2$ gauge theories
 are based on the following brane set up
in type IIA superstring theory on ${\Bbb R}^{10}$:

\begin{itemize}

\item $n$ NS 5-branes with world volumes parametrized by $x^0$, $x^1$, $x^2$,
$x^3$, $x^4$ and $x^5$; they are located at $x^7=x^8=x^9=0$ and at some
fixed value of $x^6_\alpha$ 
($\alpha=1,\dots n$), at least in classical approximation.

\item $k_\alpha$ ($\alpha=0,\dots n$)
 D 4-branes with world volumes in the $x^0$, $x^1$, $x^2$, $x^3$ and
$x^6$ directions, being suspended  in the $x^6$
direction between the $(\alpha)^{\rm th}$
and $\alpha^{\rm th}+1$ NS 5-brane. 
For $\alpha=0$ or $\alpha=n$ the D 4-branes are semi-infinite to the left 
of the first NS 5-brane or,
respectively, to the right of the $n^{\rm th}$ 
NS 5-brane\footnote{Alternatively,
within the so called elliptic models, the coordinate $x^6$ is compact,
such that $k_0=k_n$ and the correspond D 4-branes are also finite and
suspended between the first and the $n^{\rm th}$ NS 5-brane.}.
The 4-branes are located in $x^4$-$x^5$ plane at the complex coordinate 
$v=x^4+ix^5$. 

\end{itemize}

\begin{figure}[htb]
  \makebox[16cm]{
    \epsfxsize=12cm
    \epsfysize=6cm
    \epsfbox{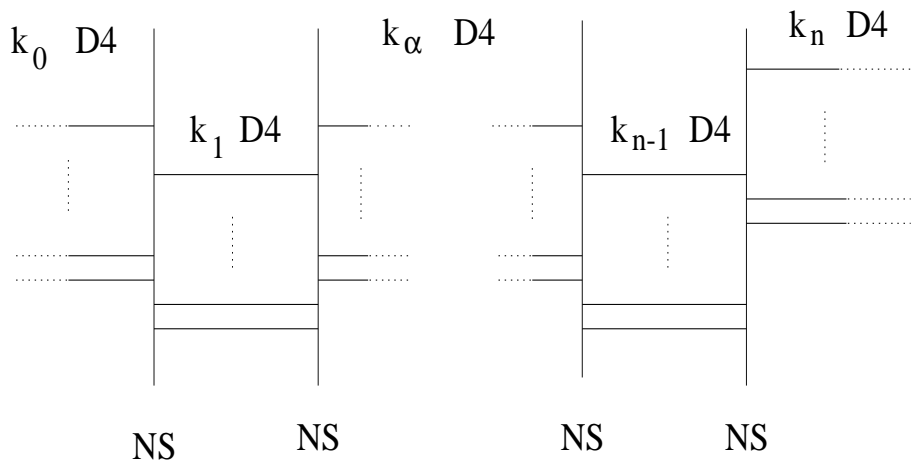}
  }
  \caption{The N=2 brane configuration}
  \label{figureBraneConfiguration}
\end{figure}

This  $N=2$ brane configuration,
which preserves 1/4 of the original supersymmetries, is summarized in figure
\ref{figureBraneConfiguration}.

%
The  four-dimensional gauge group is  given by
\begin{equation}
  G=\prod_{\alpha=1}^{n-1} SU(k_\alpha).
\end{equation}
In addition one finds
hypermultiplets in the representations:
\begin{equation} 
(k_0,\bar k_1)\oplus (k_1,\bar k_2)\oplus\dots\oplus (k_{n-1},\bar k_n).
\end{equation}
(Here, $k_0$ and $\bar k_n$ act as global flavor representations.)
  
The gauge coupling constants $g_\alpha$ of every $SU(k_\alpha)$ group factor
are determined by the differences between the positions of the NS 5-branes:
\begin{equation}
\frac{1}{g^2_\alpha}=\frac{x^6_{\alpha+1}-x^6_{\alpha}}{g_s},\label{gaugeco}
\end{equation}
where $g_s$ is the string coupling constant.

After having discussed the classical picture, let us now come to 
solution of the model after taking into account the quantum corrections.
The D 4-branes exert a force on the NS 5-branes causing them to bend.
Since the tensions of the NS 5-brane and the tensions of the D branes 
have a different behavior in terms of the string coupling constant,
$T_{NS}\sim g_s^{-2}$ compared to $T_{D}\sim g_s^{-1}$, this
bending is a quantum effect.
As a result, a NS 5-brane on which 4-branes end does not have a definite value
of $x^6$ in contrast to the classical picture. 
More specifically, when there is a bending which
moves the two NS branes towards each other, the coupling becomes
strong at high energies, i.e. we deal with an IR free theory. Conversely,
if the bending is outwards, there is an asymptotically free theory.
In four dimensions the bending is logarithmic with the distance $r$, whereas
in $d$ dimensions the local bending of the NS branes goes like
$r^{d-4}$, $r$ being the `position' of the D brane on the world volume
of the NS brane. This shows that in dimensions $d<4$ all gauge theories are
asymptotically free, and all gauge theories with finite gauge coupling
are IR free for $d>4$.
The bending is absent if the number of left D branes ending on a given
NS 5-brane is equal to the number of D branes ending from the right.

In one-loop perturbation theory in four dimensions, 
the bending of the NS 5-branes 
leads to a logarithmic
variation of $x^6_\alpha$ in terms of $v$:
\begin{equation}
   x^6_\alpha=(k_\alpha-k_{\alpha-1})\log |v|.
   \label{1loop}
\end{equation}
Inserting this into eq.(\ref{gaugeco}),  
the
logarithmic running
of the gauge coupling $g_\alpha$ is determined as
\begin{equation}
\frac{1}{g_\alpha^2}=\frac{-2k_\alpha+k_{\alpha+1}+k_{\alpha-1}}{g_s}\log |v|.
\end{equation}
Note that the prefactor $b_\alpha^{N=2}=-2k_\alpha+k_{\alpha+1}+k_{\alpha-1}$
precisely agrees with the $N=2$ $\beta$-function coefficient of the gauge
group $SU(k_\alpha)$ with $N_f=k_{\alpha+1}+k_{\alpha-1}$ fundamental matter
fields. In this way the shape of the branes incorporates
the 1-loop effects in field theory.

In $N=2$ field theory there are no higher loop effects. However there are still
non-perturbative effects due to instantons. These instantons can be seen
directly in the brane picture, namely the D 0-branes are
instantons within D 4-branes.  The problem is now
to solve the
theory by including all these effects. This can be done by ``lifting'' the
IIA configuration to M-theory \cite{witten}.
The advantage of considering the above configuration of branes in M-theory is
that the D 4-branes and NS 5-branes are in fact the same object, the M 5-brane.
The intersection of the D 4- and NS 5-branes in IIA was singular but 
this is smoothed
out in M-theory and in fact it is possible to consider all the D 4-branes and
NS 5-branes as a single M 5-brane with complicated worldvolume. However the
conditions for preserving N=2 supersymmetry restrict the embedding of the
M 5-brane worldvolume and it is possible to find the function describing this
embedding explicitly. Clearly this must incorporate the 
classical IIA brane set up as well as the field theory 1-loop corrections
through the shape of the M 5-brane. However the non-perturbative instantons
are also automatically included since D 0-branes are simply Kaluza-Klein
momentum modes of compactified M-theory. 

Let us discuss in slightly more detail how 
the M-theory embedding 
of the $N=2$ brane configurations is constructed. 
The lifting to M-theory is performed by adding the  11th  
 coordinate $x^{10}$ which is periodic with period
$2\pi R_{11}$.
Now, the complex coordinate $s_\alpha=(x^6_\alpha+ix^{10}_\alpha)/R_{11}$
describes the asymptotic positions of the M 5-branes  in the
$x^6$-$x^{10}$ plane, whereas as before $v$ denotes the asymptotic positions
of the M 5-branes in the $x^4$-$x^5$ plane (all M 5-branes have common
world volume in the 0123-directions, and are fixed at $x^7=x^8=x^9=0$).
Regarding $x^{10}$ as
$\theta$-parameter one can introduce a complex coupling constant
\begin{equation}
\tau_\alpha=\frac{\theta_\alpha}{2\pi}+i\frac{4\pi}{g^2_\alpha}=
i(s_{\alpha+1}-s_{\alpha}).
\end{equation}

Concentrating on the directions $x^4$, $x^5$, $x^6$ and $x^{10}$ we see that 
the M 5-brane world volume spans a two-dimensional
surface $\Sigma^{(2)}_{n,k_\alpha}$ in the four-manifold ${\mathbb{R}}^3\times S^1$.
A priori, $\Sigma^{(2)}_{n,k_\alpha}$ is determined by two real functions
of the coordinates $x^4$, $x^5$, $x^6$ and $x^{10}$:
\begin{eqnarray}
f_{n,k_\alpha}(x^4,x^5,x^6,x^{10})&=&0,\nonumber \\
g_{n,k_\alpha}(x^4,x^5,x^6,x^{10})&=&0.
\end{eqnarray}
However
$N=2$ space-time supersymmetry requires that
$s$ varies holomorphically with $v$, such that $\Sigma^{(2)}_{n,k_\alpha}$
is a Riemann surface
in ${\Bbb{R}}^3\times S^1$:
\begin{equation}
\Sigma^{(2)}_{n,k_\alpha}:\quad F_{n,k_\alpha}(s,v)=f_{n,k_\alpha}+
ig_{n,k_\alpha}
=0.
\end{equation} 
This means that $\Sigma^{(2)}_{n,k_\alpha}$ is a supersymmetric 
2-cycle, or in terms of \cite{gibpap}, $\Sigma^{(2)}_{n,k_\alpha}$ 
describes a
$SU(2)$ K\"ahler calibration.
The holomorphy of the equation $F_{n,k_\alpha}(s,v)$ implies that the real
functions have to obey the Cauchy-Riemann differential
equations, where $f_i$ denotes $\frac{\partial f}{\partial x_i}$:
\begin{eqnarray}
& & f_4=g_5,\qquad\;\;\;\; f_6=g_{10},\nonumber\\
& & f_5=-g_4,\qquad f_{10}=-g_6.
\end{eqnarray}

As explained in \cite{witten}, it is very useful to perform an holomorphic 
change of variables, $t=\exp(-s)$, in order to describe the asymptotic 
positions of the M 5-branes in the correct way. Specifically consider the 
complex equation
\begin{equation}
      F(t,v)=0.\label{sw}
\end{equation}
At a given value of $v$, the roots of $F$ in $t$ are the positions of the
NS 5-branes, i.e. $F$ is a polynomial of degree $n$ in $t$. On the other
hand, for fixed $t$, the roots of $F$ in $v$ are the positions of the
IIA D 4-branes.
Therefore $n$ parallel NS 5-branes
with positions at $t_i$ ($i=1,\dots ,n$) are simply
described by the function $F(t,v)=\prod_{i=1}^n(t-t_i)$, and
$k$ parallel 
D 4-branes, positioned at values $v_j$ ($j=1,\dots k$),
correspond to the choice
$F(t,v)=\prod_{j=1}^k(v-v_j)$.
Then it immediately follows that $n$ NS 5-branes intersected by $k$ D 4-branes
correspond to
\begin{equation}
F(t,v)=\prod_{i=1}^n(t-t_i)\prod_{j=1}^k(v-v_j).\label{orth}
\end{equation}
In this
case, since the number of D 4-branes to the left and to the right 
of each NS 5-brane is the same, there is no bending of the NS 5-branes by
the D 4-branes.

Next, let us briefly discuss, how the perturbative, one-loop bending can
be described \cite{witten}.
Consider the situation of one NS 5-brane with $k_0$ ($k_1$) D 4-branes 
ending on it from the left (right). Then the induced bending in the
region, where $t$ is very large, corresponds to
the following choice for $F$:
\begin{equation}
F(t,v)=v^{k_0}(t-\epsilon v^{k_1-k_0}),\label{bendtv}
\end{equation}
where $\epsilon$ is a constant.
Introducing back the variable $s$, this equation immediately leads
to eq.(\ref{1loop}), namely
the logarithmic bending of $s$ in terms of the Higgs field $v$:
\begin{equation}
s=(k_1-k_0)\log v.
\end{equation}

Putting  all these informations
from eqs.(\ref{orth},\ref{bendtv}) together,
the supersymmetric
2-cycle equation for a general $N=2$ brane configuration takes the form
\begin{equation}
\Sigma^{(2)}_{n,k_\alpha}:
\quad F_{n,k_\alpha}(t,v)=p_{k_0}(v)t^n+p_{k_1}(v)t^{n-1}+\dots 
+p_{k_{n-1}}(v)t+p_{k_n}(v),\label{n=2pol}
\end{equation}
where the $p_{k_\alpha}(v)$ are polynomials in $v$ of degree $k_\alpha$. 
The up to now unspecified parameters of the polynomials $p_{k_\alpha}(v)$
appear as the moduli of the gauge theory. 
This describes the non-perturbative solution of the model with gauge group
$\prod_{\alpha=1}^{n-1}SU(k_\alpha)$ with hypermultiplets in 
bi-fundamental representations.
For example, the pure $SU(k)$ gauge theory without matter fields, i.e
$n=2$, $k_0=k_2=0$, $k_1=k$, is described by the curve
\begin{equation}
F(t,v)=t^2+p_k(v)t+1.
\end{equation}
This is nothing else than the famous Seiberg-Witten curve of genus
$k-1$ \cite{sw,9411048,klemm,argyres}.

Why does this procedure work? The solution of the gauge theory is found by using
the duality between 11d SUGRA and IIA string theory. When
$R_{11}$ is large, 11d SUGRA is valid. Solving for the exact shape
of the M5 brane means solving the classical minimal area problem for
a tensile brane (the shape of a soap bubble) with
given boundary conditions. The requirement of supersymmetry
tells us that we should only consider special minimal area configurations,
the SUSY 2-cycles.

On the other hand we identified the gauge group and matter content at small
$R_{11}$, where we have weakly coupled IIA string theory and the analysis of
\cite{hanwit} applies. In order to decouple the bulk modes from the gauge
theory we have to take the string scale
$M_s \rightarrow \infty$  and $M_{pl} \rightarrow
\infty$ and in order to decouple the KK modes
from the finite interval, we have to also send $L \rightarrow 0$, where $L$ is
the length of the $x^6$ intervals. This all has to be done holding
$g^2_{YM}=\frac{g_s}{M_s L}$ fixed. In 11d units $g^2_{YM}=\frac{R_{11}}
{L}$. In order to keep the interacting gauge theory (fixed $g^2_{YM}$)
while decoupling the KK modes ($L\rightarrow 0$) $R_{11}$ has to go
to zero! This is the limit in which the brane setup reduces to 4d SYM. But
this is the opposite limit of the one we were able to solve, where
$R_{11}$ and hence $L$ have to be very large and we can use 11d SUGRA.
So we should only expect quantities protected by
holomorphy, like the $N=2$ prepotential,
which can't depend on the real parameter $R_{11}$, to
come out correctly. Indeed it was shown in \cite{berkeley} that
unprotected quantities like the 4-derivative terms in the
effective action disagree with field theory results.

\newpage

\section{{\it\bf N=1} Gauge Theories and Supersymmetric 3-cycles}

\subsection{$N=1$ brane boxes}
\label{sectionN1BraneBox}

Now let us discuss the brane boxes of \cite{hanzaf} which can be
used to construct $N=1$ supersymmetric gauge theories with chiral
matter content\footnote{Like for the
$N=2$ brane models, the $N=1$ brane box models can be related to
fractional branes via T-duality \cite{haur}.}.
The starting point is now a type IIB superstring with the following
branes included (see figure \ref{figureGeneralBraneBoxWeb}):

\begin{figure}[htb]
  \makebox[16cm]{
    \epsfxsize=8cm
    \epsfysize=8cm
    \epsfbox{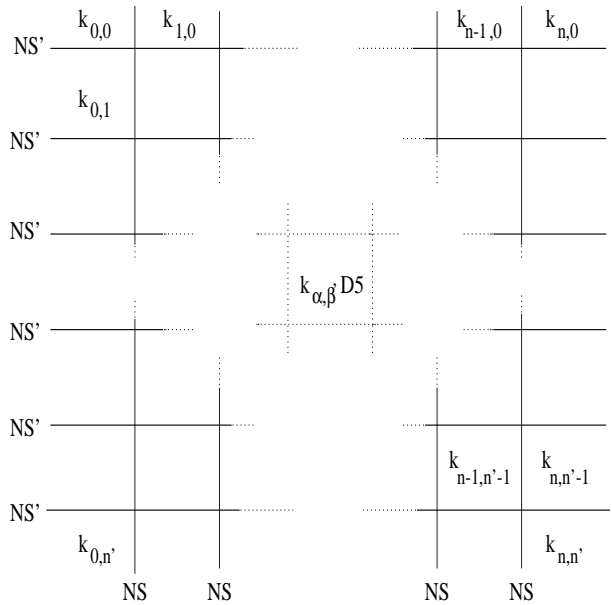}
  }
  \caption{The N=1 brane configuration}
  \label{figureGeneralBraneBoxWeb}
\end{figure}
%

\begin{itemize}

\item

$n$ parallel NS 5-branes with world volumes along the (012345)-directions.
These 5-branes are fixed at $x^7=x^8=x^9=0$ and are placed at arbitrary
positions in $x^6$.

\item

$n'$ parallel NS' 5-branes with world volumes along the (012367)-directions.
The NS' branes are fixed at $x^5=x^8=x^9=0$ and are placed at arbitrary
positions in $x^4$.

\item
D 5-branes with world volumes along the (012346)-directions. The D 5-branes
can take different position on the NS 5-branes in the $x^5$-direction and also
different positions on the NS' 5-branes in the $x^7$-direction.
Depending on the specific model one likes to discuss, the directions
$x^4$ and $x^6$ can be either uncompactified or periodic 
(elliptic models). We will concentrate on the non-elliptic
models. It follows that
the D 5-branes are finite in the 
directions $x^6$, $x^4$ in case 
they are placed inside the `inner' boxes. However
they are semi-infinite in case
they end  only on one NS (NS') brane (`outer' boxes). 

\end{itemize}

This brane configuration preserves 1/8 of the original supersymmetry.
We see that a generic configurations consists of a grid of 
$(n+1)(n'+1)$ boxes built
by
$n$ NS 5-branes branes and $n'$ NS' 5-branes in the $x^4$-$x^6$ plane.
We are labelling the boxes by the two indices 
$\alpha,\alpha'$ where $\alpha=0,\dots , n$ and
$\alpha'=0,\dots , n'$. The  $(n-1)(n'-1)$
`inner' boxes with $\alpha=1,\dots, n-1$, $\alpha'=1,\dots ,n'-1$
have always finite area whereas the remaining `outer' boxes have infinite
size in case of uncompactified directions $x^4$ and $x^6$.

Now, let $k_{\alpha,\alpha'}$ denote the number of D 5-branes which are placed
in the box $\lbrack\alpha,\alpha'\rbrack$.
The suspended D 5-branes inside the inner
boxes give rise to the following gauge group
in four dimensions:
\begin{equation}
G=\prod_{\alpha=1}^{n-1}\prod_{\alpha'=1}^{n'-1}SU(k_{\alpha,\alpha'}).
\end{equation}
The associated classical gauge coupling constants are given by the 
area of the corresponding box $\lbrack\alpha,\alpha'\rbrack$:
\begin{equation}
\frac{1}{g_{\alpha,\alpha'}^2}=
\frac{(x^4_{\alpha+1}-x^4_\alpha)(x^6_{\alpha'+1}-x^6_{\alpha'})}{g_s}.
\label{n=1gaugec}
\end{equation}

The matter content of the model consists of three types of chiral $N=1$
representations. First they are `horizontal' chiral bi-fundamentals
$H_{\alpha,\alpha'}$ 
in the representations $(k_{\alpha,\alpha'},\bar k_{\alpha+1,\alpha'})$
of $SU(k_{\alpha,\alpha'})\times SU(k_{\alpha+1,\alpha'})$. 
Second there
exist `vertical' chiral bi-fundamentals $V_{\alpha,\alpha'}$  in the
representations $(k_{\alpha,\alpha'},\bar k_{\alpha,\alpha'+1})$ of
$SU(k_{\alpha,\alpha'})\times SU(k_{\alpha,\alpha'+1})$;
finally we have
`diagonal' chiral fields $D_{\alpha,\alpha'}$ in the representations
$(k_{\alpha,\alpha'},\bar k_{\alpha-1,\alpha'-1})$ 
of $SU(k_{\alpha,\alpha'})\times SU(k_{\alpha-1,\alpha'-1})$ 
($\alpha,\alpha'>1$).
In this context the groups $SU(k_{\alpha,\alpha'})$ with $\alpha=0,n$ 
or $\alpha'=0,n'$
act as global flavor symmetries if the directions $x^4$ and $x^6$ are
uncompactified. Note that the choices for the $k_{\alpha,\alpha'}$ are 
 severely constrained by the requirement of absence of anomalies.
If all three types of matter multiplets are present
then there exists a classical  superpotential of the
following form:
\begin{equation}
W=\sum_{\alpha,\alpha'}H_{\alpha,\alpha'}V_{\alpha+1,\alpha'}
D_{\alpha+1,\alpha'+1}-\sum_{\alpha,\alpha'}H_{\alpha,\alpha'+1}
V_{\alpha,\alpha'}D_{\alpha+1,\alpha'+1}. \label{supo}
\end{equation}

One of the  simplest (non-elliptic)
models  is given by the choice
$n=n'=2$, $k_{1,1}=N_c$ and $k_{0,1}=k_{2,1}=N_f$, 
whereas $k_{\alpha,\alpha'}=0$ for
$\alpha'=0,2$.
This choice of brane boxes
corresponds to the supersymmetric QCD with $G=SU(N_c)$ and with $N_f$ 
fundamental plus
antifundamental chiral fields.
A second way to obtain SUSY QCD with $N_f$ fundamental plus antifundamental
matter fields is given by the choice $k_{1,1}=N_c$ and $k_{0,1}=k_{2,1}=
k_{1,0}=k_{1,2}=N_f/2$ and zero otherwise. 
Finally the same spectrum can by realized by the
choice $k_{1,1}=N_c$ and $k_{\alpha,\alpha'}=N_f/3$ ($\alpha,\alpha'\neq1$).
However in this case a superpotential of the type eq.(\ref{supo}) is present.

So far we have only discussed the classical field theory. Of course it is
essential to understand the quantum features as well. Chiral $N=1$
exhibit a huge variety of interesting quantum phenomena. Especially
the generic theory will have an anomaly which should show up as
an inconsistency of the brane box as a string background. In these general
cases the
bending of the brane boxes isn't well understood yet, but some special
cases can be analyzed.

It is clear that the bending of the NS and NS' branes depends on the number
$k_{\alpha,\alpha'}$ of D 5-branes in each box.
A very special class of $N=1$ gauge theories is given by the
{\sl finite} models for which all $\beta$-functions and all anomalous
dimensions vanish
to all orders in perturbation theory \cite{leighstrassler,HaStrUr}.
This condition includes the vanishing of the one-loop $\beta$-functions.
In the brane picture complete finiteness means that
all NS and NS' branes do not bend at all, i.e.
if the number of D 5-branes in every box is the same \cite{HaStrUr}.
Then obviously, $N_f=3N_c$ for every gauge group factor, and the
one-loop $\beta$-functions are zero.

The corresponding brane setup consists of several
branes put on top of each other.
Each of the branes preserves 1/2 of the supersymmetries, together they still
preserve 1/8, so the intersection is BPS. This ensures that the static branes
don't exert any force on each other. We can freely move the constituent
branes since they don't feel the presence of the other branes at all.
Motions of the branes in the 46 plane corresponds to changing the
areas of the various boxes and hence to changing the gauge couplings.
Taking NS (NS') branes away along the $x^7$ ($x^5$) direction
destroys the box structure and
corresponds to turning on FI terms.

\begin{center}
\begin{figure}[htb]
  \makebox[16cm]{
    \epsfxsize=11cm
    \epsfysize=4cm
    \epsfbox{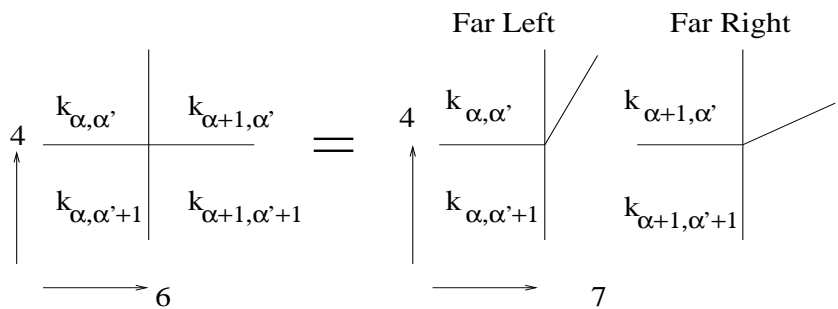}
  }
  \caption{Basic cross configuration}
  \label{figureUniformBasic}
\end{figure}
\end{center}

Another special situation is that of uniform bending. The condition
of uniform bending was first introduced by \cite{gimongremm}. There
it was argued to be necessary for consistency. As we will see this
is too stringent. However uniform bent setups are very special and
allow for a precise treatment of the quantum properties. To motivate
the uniform bending requirement consider the basic cross configuration
of figure \ref{figureUniformBasic}.

%

For $x^6 \rightarrow - \infty$ (to the far left) the effects of the
NS brane on the bending of the NS' brane should be negligible.
The D5 ending on the NS' brane just looks like a 5d gauge theory
with 8 supercharges and leads to the standard linear bending
\cite{aharonyhanany}. The slope of the bending is given by the
difference of branes ending from either side, hence
\begin{equation}
\mbox{slope}_{x^6 \rightarrow - \infty} = k_{\alpha,\alpha'} 
                                        - k_{\alpha,\alpha'+1}.
\end{equation}
For the same reason we will have linear bending to the far right, that
is for  $x^6 \rightarrow  \infty$, the slope this time given
by
\begin{equation}
\mbox{slope}_{x^6 \rightarrow  \infty} = k_{\alpha+1,\alpha'} 
                                       - k_{\alpha+1,\alpha'+1}.
\end{equation}

The observation of \cite{gimongremm} was that if
\begin{equation}
\label{uniform}
k_{\alpha,\alpha'} - k_{\alpha,\alpha'+1} =   k_{\alpha+1,\alpha'} 
                                          - k_{\alpha+1,\alpha'+1}
\end{equation}
the bending on the far left is the same as on the far right
and one may expect that the shape of the NS' in fact does not change
at all as a function of $x^6$. In \cite{haur} it was observed that the most
general setup compatible with condition (\ref{uniform}) can
be achieved by ``sewing'' together $N=2$ models, that is we take
branes corresponding to 5d gauge theory built out of NS and D5 branes
and move them on top of a similar setup build out of NS' and D5 
branes\footnote{In \cite{haur} they also allowed sewing in a third kind
of $N=2$ system connected to diagonal lines in the box setup. This
doesn't lead to uniform bent models anymore and should hence be treated
separately.}, as illustrated in figure \ref{figureUniformBraneBoxWeb}.

\begin{center}
\parbox{14.7cm}
{
\refstepcounter{figure}
\label{figureUniformBraneBoxWeb}
\font\bildfont=cmcsc10
\parbox{4.7cm}
{
  \makebox[4.5cm]
  {
     \epsfxsize=4.5cm
     \epsfysize=2.25cm
     \epsfbox{bilder/braneconf.eps}
  }
}\hfill
\parbox{4cm}
{
  \makebox[3.5cm]
  {
     \epsfxsize=3.5cm
     \epsfysize=4.5cm
     \epsfbox{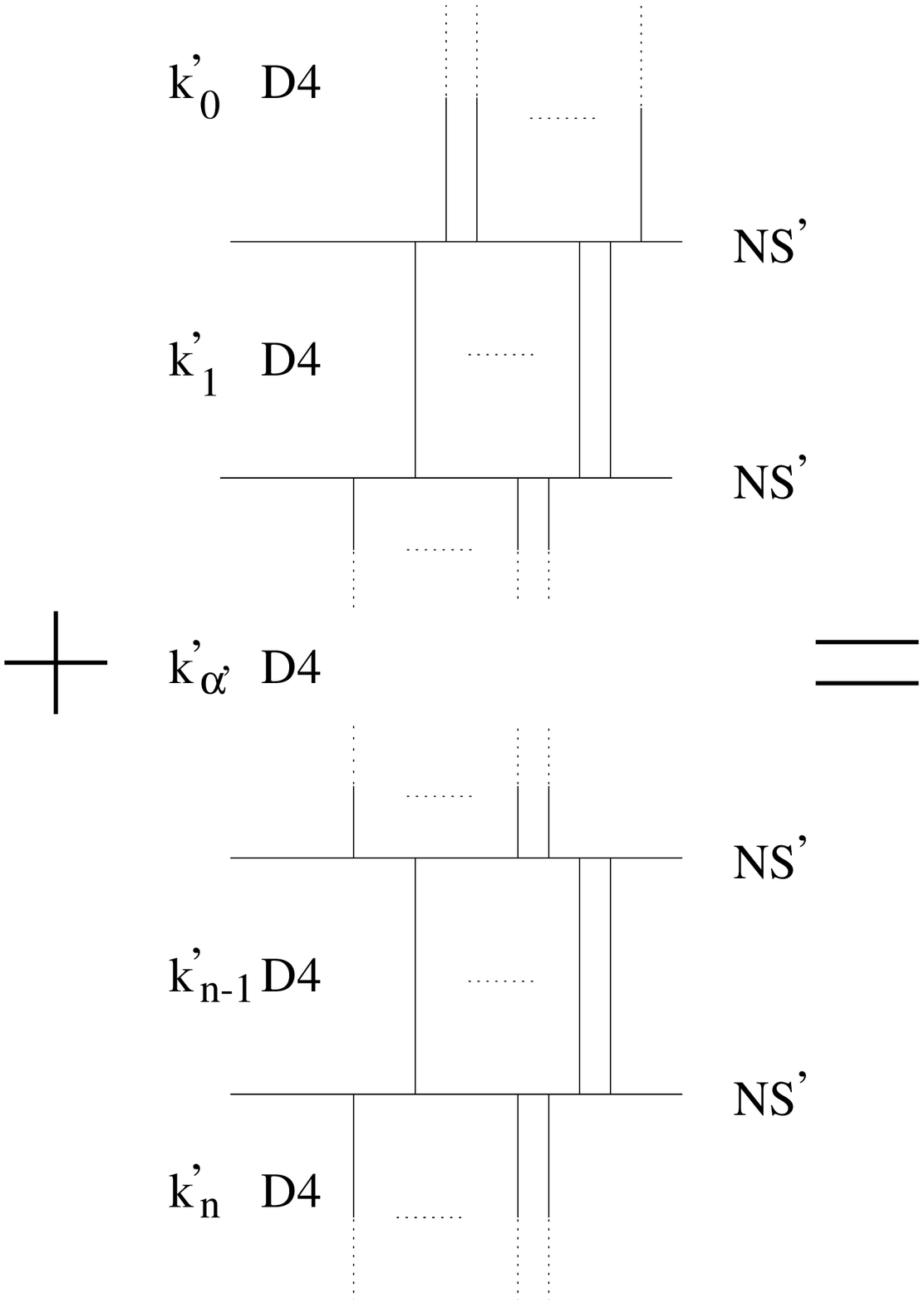}
  }
}\hfill
\parbox{6cm}
{
  \makebox[6cm]
  {
     \epsfxsize=6cm
     \epsfysize=6cm
     \epsfbox{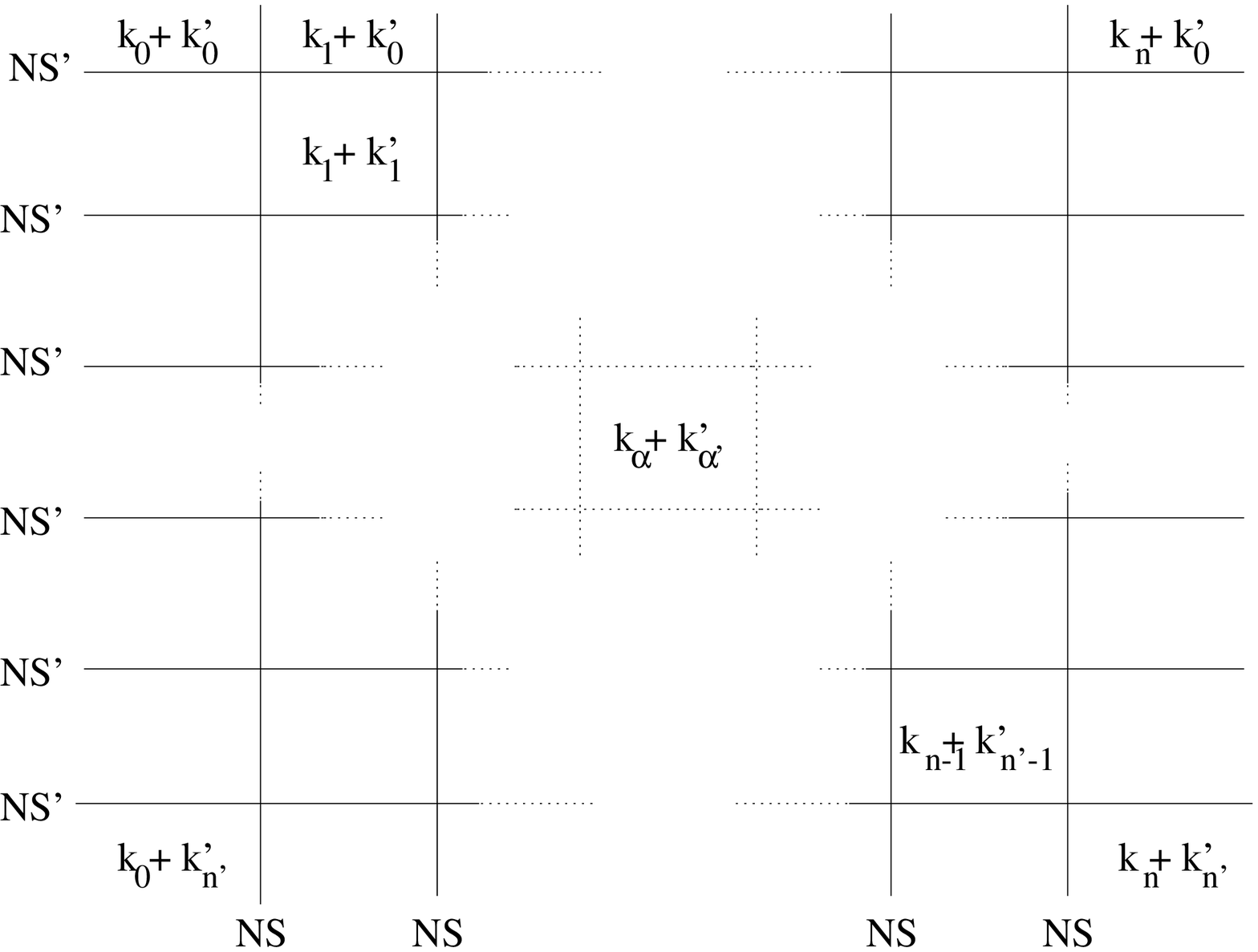}
  }
}\hfill\\
 \center{{\bildfont\large Figure} \thefigure $:\;$ Sewing of N=2 
          models.} 
}
\end{center}
%

This sewing can be taken quite literally: as in the finite case there is
a no force condition between the constituent pieces, since their intersection
still preserves some supersymmetry. So we are free to move them independently.
These deformations should correspond to marginal couplings in the
field theory. Since we are not just tuning the distance between
two NS branes but are actually moving around compound systems, these
marginal operators won't just be the gauge coupling as in the finite case
but will also involve the superpotential couplings. Using the methods
of \cite{leighstrassler}, \cite{HaStrUr} were indeed able to show that
the field theory has these marginal operators if the conditions
(\ref{uniform}) are satisfied for all boxes. Since the subsystems
don't influence each other, the exact bending is given in terms of the linear
bending of the subsystems. All uniform bent systems are anomaly free.

The question remains, how to decide whether the more general setups are
anomaly free. So far we were only able to discuss some very special cases,
e.g. not including pure SYM. 
Some progress in understanding the bending in these cases
has been made in \cite{randall}, however without reaching
a final answer. Several aspects of this problem can be easier understood in
a T-dual picture. Consider an elliptic model, that is we take the
$x^4 - x^6$ plane to be compactified on a torus. For a finite
model, T-dualizing
these two compact directions turns the D5 branes into D3 branes on an 
${\Bbb{C}}^3/\Gamma$ orbifold background. Following
\cite{LKS} it was argued in \cite{haur} that the generic box T-dualizes
into an orbifold with fractional branes, that is where the orbifold
group is embedded into the gauge group via some other representation than
the regular one. On the orbifold side one immediately faces the issue of
tadpole cancellations. Non-vanishing tadpoles signal the presence of a source,
that is a net charge sitting in the internal space. So in orbifold
compactifications the tadpoles have to vanish for consistency. Since
we are dealing with non-compact backgrounds some non-vanishing tadpoles
may be tolerated. The relevant space is the FP set of the orbifold action
transverse to the D3 brane. For the $N=2$ case (a D3 brane in 0123 with
an orbifold acting on 6789 space) the FP set is 2 dimensional. A net charge
in 2d will give rise to a logarithmic singularity. In \cite{LKS} it was
shown that this divergence is nothing else but the running of the
gauge coupling. Tadpole cancellation is hence not required for
consistency. Vanishing of the tadpoles is equivalent to finiteness of
the gauge theory.

In the same spirit Leigh and Rozali analyzed the orbifold dual of the
brane boxes \cite{leighrozali}. A generic orbifold element will
leave a 0d FP set. These tadpoles have to vanish. Otherwise
the gauge theory is anomalous. However some orbifold elements will
leave a 2d FP set untouched. The corresponding tadpoles will
only lead to logarithmic divergences which once more can be identified
as the running gauge coupling. So vanishing of all tadpoles
again implies finiteness of the gauge theory. Vanishing of the tadpoles
for the 0d FP is required for anomaly freedom. Leigh and Rozali indeed
showed that for all brane boxes leading to anomaly free gauge theories
these 0d tadpoles vanish.

\subsection{T-duality, M-theory embedding and the emergence of the 3-cycles}

Now let us describe the the strong coupling limit of the $N=1$ via
embedding the brane boxes into M-theory. Since our original
brane configurations is in the type IIB
string, we have first to perform a T-duality transformation
to the type IIA superstring before we can perform the M-theory
embedding.
We do not want to touch the NS and NS' 5-branes, and 
we also do not want to create any D6-branes; therefore
we will T-dualize over one
of the spatial directions common to all branes. To be specific we now
assume that $x^3$ is periodic with radius $R_3^{IIB}$ 
and we perform the T-duality
with respect to the $x^3$-direction. This leads to the following brane
configuration: 

\begin{itemize}

\item

$n$ parallel NS 5-branes with world volumes along the (012345)-directions.
These 5-branes are fixed at $x^7=x^8=x^9=0$ and are placed at arbitrary
positions in $x^6$.

\item

$n'$ parallel NS' 5-branes with world volumes along the (012367)-directions.
The NS' branes are fixed at $x^5=x^8=x^9=0$ and are placed at arbitrary
positions in $x^4$.

\item
D 4-branes with world volumes along the (01246)-directions.
These D 4-branes take different $x^5$ positions on the NS 5-branes and also
different $x^7$ positions on the NS' 5-branes.
In addition the D 4-branes can have 
arbitrary positions in the compactified spatial
$x^3$-direction.

\end{itemize}

This configuration 
preserves like before 1/8 of the original supersymmetries and
corresponds to a three-dimensional gauge theory with
$N=2$ space-time supersymmetry. The three-dimensional 
gauge theory can be simply obtained
from the four-dimensional $N=1$ models by circle compactification on $S^1$
in the $x^3$-direction. 
In the decompactification limit, $R_3^{IIB}\rightarrow\infty$,
the four-dimensional $N=1$ gauge theories are rediscovered. 
On the hand, for $R_3^{IIB}\rightarrow 0$, the theory
is truly three-dimensional. Note that in three
dimensions, a new Coulomb branch can be opened, since the three-dimensional
vector multiplets contain one real scalar degree of freedom. The corresponding
modulus $v$ is associated in the brane picture with the positions of the
D 4-branes in the $x^3$-direction. 

The three-dimensional gauge coupling is classically related to the 4d
gauge coupling as $1/g_3^2=R_3^{IIB}/g_4^2$.
So in the limit $R_3^{IIB}\rightarrow\infty$
one must send $g_3\rightarrow 0$ in order to have a finite coupling $g_4$.
The scalar field in the vector multiplet live on a dual circle with
radius $R_3^{IIA}=1/R_3^{IIB}$. So in the 4d limit, $R_3^{IIA}\rightarrow 0$,
one has to integrate out the fields 
with masses of order $v$ corresponding to the Coulomb branch.
In this way we can regard $v$ as the parameter 
which sets the scale $\Lambda$ of the four-dimensional gauge theory.
So in order to determine the logarithmic `running' of the four-dimensional
gauge coupling $g_4$ in terms of $\Lambda$, 
$1/g_4^2=b^{N=1}\log\Lambda$, we will be in particular interested
in the bending of the coordinates $x^4$ and $x^6$ in terms of $x^3$.
(This precisely corresponds to the logarithmic running of 
$x^6$ in terms of the Higgs field vev $v=x^4+ix^5$ in case of the $N=2$
brane models.) This will be further discussed in section 
\ref{sectionUniformBending}.
Note that in the 3-dimensional limit, $R_3^{IIB}\rightarrow 0$,
the pure Yang-Mills gauge theory has no stable supersymmetric
groundstate unlike the 4d theory. Many more details of the dynamics and
superpotentials of  three-dimensional, $N=2$ supersymmetric gauge theories can be 
found in \cite{3dgauge}.

After the duality in $x^3$ we are now ready for
lifting the above configuration to M-theory by adding the
period direction $x^{10}$ with radius $R_{11}$.

Then the intersection of all branes is described by the smooth configurations
of M 5-branes.
Like in the $N=2$ case the singular intersections of the NS, NS' and D 4-branes
are described in M-theory by a single smooth M 5-brane. Asymptotically,
this M 5-brane takes the shape of the classical IIA branes:

\begin{itemize}

\item
The NS 5-branes asymptotically correspond to  M 5-branes
which extend in the (012345)-space and take different positions in the
$x^6$, $x^7$, $x^8$, $x^9$ and $x^{10}$ directions.

\item
The NS' 5-branes asymptotically look like M 5-branes with world volumes
along the (012367)-space and positions in $x^4$, $x^5$, $x^8$,
$x^9$ and $x^{10}$. 

\item
Finally, the 
asymptotics of the D 4-branes is given by M 5-branes with
world volumes in $x^0$, $x^1$, $x^2$, $x^4$, $x^6$ and $x^{10}$ and
positions in $x^3$, $x^5$, $x^7$, $x^8$ and $x^9$.

\end{itemize}

So all branes have common world volumes in the (012)-space and are all located
at $x^8=x^9=0$.
Therefore,
to characterize the M-theory configurations we have to focus on the
six-dimensional space spanned by the
coordinates $x^3$, $x^4$, $x^5$, $x^6$, $x^7$ and $x^{10}$. Each 
asymptotic brane fills three particular directions in
this space.
This means that the general embedding of the M-theory  5-branes is described by
a three-dimensional submanifold in ${\mathbb{R}}^4\times T^2$.
Supersymmetry requires this to be a supersymmetric 3-cycle.
In the language of \cite{gibpap} this is called a $SU(3)$ special
Lagrangian calibration (SLAG) which breaks 7/8 of the supersymmetries.

Let us again briefly consider the validity of our approach.
In 11d units the 3d gauge coupling is given by $g^2_{YM}=\frac{R_{11}}
{A}$ where $A$ is the area of the box. As in the $N=2$ case there
are two distinct limits if we want to
keep $g^2_{YM}$ fixed: for small $R_{11}$ and $A$ the KK modes
decouple and the brane setup reproduces the gauge theory. For large
$R_{11}$ we can solve using 11d SUGRA, that is by solving for
the SUSY 3-cycle. So we are again only solving MQCD and
not really the gauge theory. However, as for $N=2$, all
holomorphic quantities should be encoded in the geometry
of the 3-cycle. That is of the terms in the 2-derivative
approximation of the effective action, the holomorphic
gauge coupling and the superpotential should be encoded in
the 3-cycle, whereas the K\"ahler potential probably escapes
our control.

\subsection{Brane Cubes and M-theory}

As already mentioned in the original work of \cite{hanzaf} the
idea of brane boxes can naturally be extended to brane cubes and brane
hypercubes. Each time we add one more NS brane with yet another orientation
breaking another half of the supersymmetry. The D brane ends on these
new NS branes as well, so that the brane spans a 3d cube or 4d hypercube
instead of the 2d box we considered so far. Let us briefly discuss,
how these configurations are lifted to M-theory. We will find that
the situation is especially interesting in the case of brane cubes, where
one can find two distinct models, one with chiral and one with non-chiral
SUSY. 

For simplicity let us consider the brane cubes directly on the IIA
side with the D4 brane suspended between the NS branes. This is the
setup that lifts to M-theory in a straight forward fashion. The third
NS brane that we add should have a fixed $x^2$ position, so that
the D4 brane is finite in this direction as well as in $x^4$ and $x^6$.
There are two distinct possibilities to do so. The first is to add
an NS'' brane along 014567. This is the setup considered in \cite{9806177}.
It leads to a chiral $N=(2,0)$ supersymmetric gauge theory in $d=2$.
Rotations in 89 space give rise to the $U(1)_R$ symmetry. The dual
orbifold consists of D1 branes living on top of an 
${\mathbb{C}}^4/ \Gamma$ orbifold, where $\Gamma$ is a subgroup
of $SU(4)$. By the same reasoning as for the brane box we find that
this chiral brane cube should lift to M-theory via a SUSY 4-cycle
in the 7d 234567 space, that is via a SUSY 4-cycle associated with $G_2$
holonomy.

The second possibility is to have the NS'' brane in 012468. This leaves
us with $N=(1,1)$ in 2 dimensions. Since this time all three types of
NS branes have a common direction ($x^3$) we can perform a T-duality
to type IIB as for the box, leading to a 3d $N=1$ theory. This
time the dual orbifold is a $G_2$ orbifold while the lift to M-theory
has to be performed via an $SU(4)$ SLAG 4-cycle in ${\mathbb{C}}^4$.
Therefor this non-chiral cube requires the same techniques as the
SLAG 3-cycle. By simply adding another NS along 23469
we find the brane hypercube and its lift via an $SU(5)$ SLAG.

\subsection{The supersymmetric $d$-cycles}

\subsubsection{The $d$-cycle equations}
\label{SectiondCycleEquation}

\begin{figure}[htb]
\makebox[16cm]{
 \epsfxsize=6cm
 \epsfysize=6cm
 \epsfbox{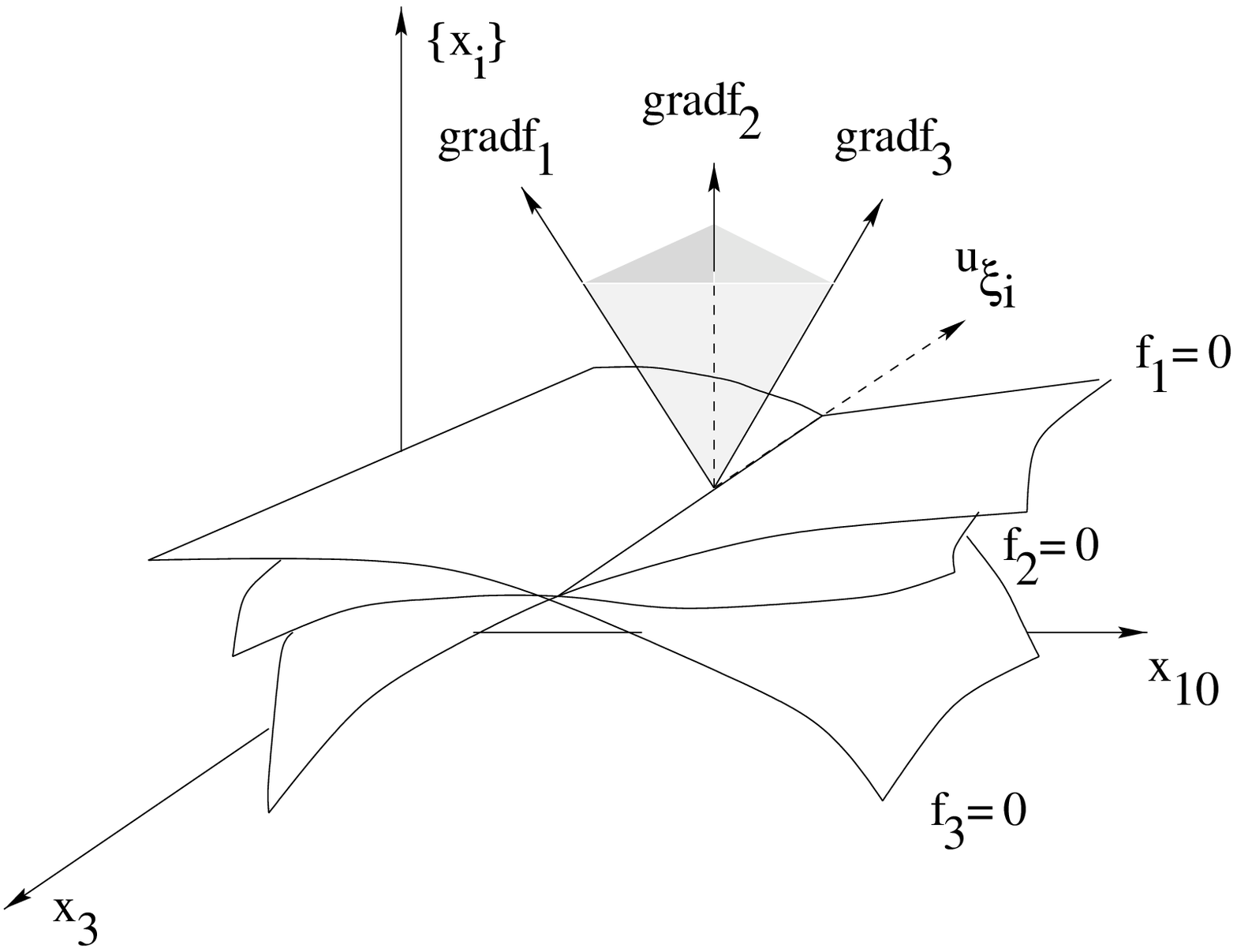}
}
\caption{The intersecting configuration}
\label{figureIntersection}
\end{figure}

A d-dimensional `curve' $\Sigma^{(d)}$, embedded into $2d$-dimensional 
flat space ${\mathbb{R}}^{2d}$ with coordinates $x^i$ ($i=1,{\ldots\,},2d$) 
can be described at least locally by the zero locus of $d$ real functions 
$f^m(x^1,{\ldots\,},x^{2d})$:
\begin{equation}
\Sigma^{(d)} = {\Bbb{V}}(f^1,{\ldots\,},f^d) = \{(x_1,{\ldots\,},x_{2d})\,|\,
                 f^m(x^1,{\ldots\,},x^{2d})=0,\;  m=1,{\ldots\,},d\}.
                 \nonumber
\end{equation}
If one wants to deal with a so called supersymmetric $d$-cycle, the choice
of the functions $f^m$ is highly constrained.
To study these restrictions  we first introduce $d$ real coordinates
$\xi_i$ 
($i=1,\dots ,d$) 
which parametrize the curve $\Sigma^{(d)}$. Furthermore we consider
complex coordinates $u^i$, $u^i=x^{2i-1}+ix^{2i}$, of ${\mathbb{C}}^d$.
Then the $d$-cycle can be characterized by making the complex $u^i$ to be 
functions of the real coordinates $\xi_i$, i.e. by the following embedding 
map $i$ from $\Sigma^{(d)}$ into ${\mathbb{C}}^d$:
\begin{equation}
  i:\Sigma^{(d)}\longrightarrow {\mathbb{C}}^d:
    \qquad \xi_i \longrightarrow u^i(\xi_i),\quad i=1,\ldots, d.
\end{equation} 
The intersection configuration (for the case $d=3$) is depicted in figure
\ref{figureIntersection}.
%
Now by applying the partial derivative $\partial_{\xi_k}$ to the defining 
equations $f^m$ of the $d$-cycle, we get the following relations: 
\begin{equation}
  \sum_{n=1}^d (f^m_{u^n} u^n_{\xi_k} + f^m_{\bar{u}^n}
  \bar{u}^n_{\xi_k}) =0
\end{equation}
(here $f^m_{u^n}=\frac{\partial f^m}{\partial u^n}$, $u^n_{\xi_k}=\frac{\partial u^n}{
\partial\xi_k}$).
These can be grouped into the following matrix expressions
\begin{eqnarray}
\left(\begin{array}{cccc} 
        f^1_{u^1} &  f^1_{u^2} & \dots & f^1_{u^d}\\
        f^2_{u^1} &  f^2_{u^2} & \dots & f^2_{u^d} \\
        \dots &\dots &\dots &\dots \\
        f^d_{u^1} &  f^d_{u^2} & \dots &f^d_{u^d} \\
      \end{array}
\right)\cdot 
\left(\begin{array}{c}
         u^1_{\xi_k} \\
         u^2_{\xi_k} \\
         \dots \\
         u^d_{\xi_i} \\
      \end{array}
\right) &=& (-1)^{d}
\left(\begin{array}{cccc} 
        f^1_{\bar{u}^1} &  f^1_{\bar{u}^2} & \dots &f^1_{\bar{u}^d} \\
        f^2_{\bar{u}^1} &  f^2_{\bar{u}^2} & \dots &f^2_{\bar{u}^d} \\
        \dots & \dots  &\dots &\dots \\
        f^d_{\bar{u}^1} &  f^d_{\bar{u}^2} & \dots &f^d_{\bar{u}^d} \\
      \end{array}
\right)\cdot 
\left(\begin{array}{c}
         \bar{u}^1_{\xi_k} \\
         \bar{u}^2_{\xi_k} \\
         \dots \\
         \bar{u}^d_{\xi_k} \\
      \end{array}
\right)\nonumber
\end{eqnarray}
We will denote  the left matrix by $M$ and  the 
right matrix  as $\bar{M}$, henceforth. 
Note, the sign in front of $\bar{M}$ depends 
on the dimension $d$ of the cycle. With the help
of these matrices we can express the bared derivatives 
by the unbared ones in the following way:  
\begin{equation}\
    \partial_k \bar{U} = (-1)^d \bar{M}^{-1}M\cdot\partial_k U 
                       = N\cdot\partial_k U .\label{lemma}
\end{equation}
By definition  $N$ shares the properties:
\begin{enumerate}
\item $N^{-1} = (-1)^d M^{-1}\bar{M} = \bar{N}$
\item $\left|\det \, N\right| = 1$
\item If $\;N=N^T\;\;\;\Rightarrow N\in U(d)$
\item If further $\;\det \, N=1\;\;\Rightarrow N\in SU(d)$
\end{enumerate}

Remembering the $d$-cycle should be supersymmetric we can 
ask for restrictions of the matrix $N$ following from this condition. It is 
well known that the notion of supersymmetric cycles \cite{BBS} coincides with 
the notion of special Lagrangian submanifolds \cite{HL}\footnote{see the 
footnote on page \pageref{Tragik}}
which can be rephrased in terms of the embedding map 
$i$: $d$-cycle $\longrightarrow {\mathbb{C}}^d$ and the two conditions:
\begin{eqnarray}
      i^\ast \MyIm \Omega &=& 0  \;\;\;\;  {\rm volume\;\; minimizing}
      \nonumber\\
      i^\ast \omega &=& 0      \;\;\;\;  {\rm Lagrangian\;\; submanifold}
\label{volmin}
\end{eqnarray}
Here $\Omega$ is the complex structure of
${\mathbb{C}}^d$, which we can choose to be 
$\Omega=du^1\wedge du^2\wedge\dots\wedge du^d$;
$\omega$ is the canonical K\"ahler form, 
$\omega = \frac{1}{2i}\sum_{i} du^i\wedge d\bar{u}^i$.
As it is shown in appendix \ref{AppendixDerivationOfTheCycleEquation} 
for the case of $d$-cycles, from the first equation we derive straightforward
that $N$ restricted to the $d$-cycle must be of unit determinant. 
\begin{eqnarray}
    \det \, N\,|_{{\Bbb{V}}(f^1,{\ldots\,},f^d)} & = & 1
\end{eqnarray}
But then to utilize this for computation in the embedding space we 
reformulate that condition a little bit. If 
\begin{eqnarray}
          I({\Bbb{V}}) = \{f\in C^{\infty}({\Bbb{R}}^{2d})\,|\,
                           f|_{\Bbb{V}} = 0 \} \nonumber
\end{eqnarray}
denotes the ideal of functions vanishing on ${\Bbb{V}}(f^1,{\ldots\,},f^d)$,
the above equation can be rewritten as
\begin{eqnarray}
    \det \, N & = & 1 + \left[\,{\rm some}\; g\in I({\Bbb{V}})\,\right] 
    \nonumber
\end{eqnarray}
This kind of non uniqueness is apparent through out the equations. To get 
a handle for that is the main obstruction for concrete computations. 

To keep that difference in mind but deal with the equations as there is no 
difference at the same time, we replace the equality sign by the congruence 
symbol ($\equiv$), i.e.
\begin{eqnarray}
    \det \, N & \equiv & 1.\label{detcond}
\end{eqnarray}

By close inspection of the second equation in (\ref{volmin}) 
(see again appendix 
\ref{AppendixDerivationOfTheCycleEquation})
one is led to a further 
restriction on $N$, namely
\begin{equation}
N\equiv N^T.\label{ntcond}
\end{equation}
So, in this way, we have translated the conditions of having a supersymmetric
cycle to restrictions on our defining equations $f^m$.

In summary, all what we have done so far can be formulated 
in a short but important 
proposition which is the starting point for all further 
computations:\\[10pt]
{\bf Proposition: }{\it A $d$-cycle, represented as an intersection of 
             $d$ real  valued functions is\linebreak supersymmetric, 
             iff 
             $N\equiv N^{T}$ and $\det \, N\equiv 1$.}\\[10pt]
It will turn out to be very useful to reformulate the last proposition
$N\equiv N^{T}$ in a different, but equivalent way.
Namely, it is not difficult to show that the requirement $N\equiv N^T$ is 
equivalent to the condition that the matrix $MM^+$ should be real modulo 
$I({\Bbb{V}})$.
To prepare this reformulation we remark that by the split of the coordinates 
of ${\Bbb{R}}^{2d}$ into the coordinates of ${\mathbb{C}}^d$ they inherit an 
intrinsic meaning as the spatial and momentum variables of symplectic
geometry. This is given by  
\begin{equation}
    u^i=q^i+ip^i,
\end{equation} 
i.e. the real part of $u^i$ gets the meaning of a spatial coordinate whereas 
the $p^i$ is a momentum variable. Then we are free to define the convenient 
Poisson brackets of phase-space functions $\{ f,g\}$. This is done  in 
the standard way as
\begin{equation}
\{ f,g\} =  \sum_{i=1}^d
            \left(
               \frac{\partial f}{\partial q^i}\frac{\partial g}{\partial p^i}
              -\frac{\partial f}{\partial p^i}\frac{\partial g}{\partial q^i}
	    \right) 
         = \sum_{i=1}^d (f_{2i-1}g_{2i}-f_{2i}g_{2i-1}),
\end{equation}
where $f_{2i-1}=\frac{\partial f}{
\partial q^i}=\frac{\partial f}{\partial x^{2i-1}}$ and
$f_{2i}=\frac{\partial f}{\partial p^i}=\frac{\partial f}{\partial x^{2i}}$.
Then the matrix
$MM^+$ reads
\begin{eqnarray}
 (MM^+)_{mn} =\,<\nabla\,f^m,\nabla\,f^n\,>\pm\, i\cdot\{ f^m,f^n\} .
\end{eqnarray}
So $MM^+$ is a real matrix modulo $I({\Bbb{V}})$, i.e. $N\equiv N^T$, if all 
Poisson brackets among the defining functions $f^m$ and  $f^n$ vanish:
\begin{equation}
\{ f^m,f^n\}\equiv 0.\label{poisson}
\end{equation}
So we get a more suitable set of equations for concrete calculations. On 
the other side the last equations can be understood very natural (see 
appendix \ref{AppendixLiouville}).\\[10pt]
{\bf Corollary: }{\it A $d$-cycle, represented as an intersection of 
             $d$ real  valued functions is\linebreak supersymmetric, 
             iff 
             $\{f^i,f^j\}\equiv 0$ and $\det \, 
             N\equiv 1$\label{Tragik}\footnote{After finishing and 
                                             submitting our paper we 
                                             became aware of \cite{HL} 
                                             where it was already 
                                             stated, however without 
                                             detailed proofs, that a 
                                             special Lagrangian submanifold
                                             is determined by the eqs. 
                                             (\ref{detcond}, \ref{poisson})}.}

\subsubsection{Supersymmetric 2-cycles}

Now we want to rederive the known result for the case of supersymmetric 
two-cycles to give a simple check of our formalism and to establish 
our point of view on the meaning of the defining equations of 
the $d$-cycle.
If we look at the brane configuration, 
\begin{table}[htb]
\renewcommand{\arraystretch}{1}
\begin{center}
\begin{tabular}{|c|ccccccccccc|}
\hline
NS : & 0 & 1 & 2 & 3 & 4 & 5 &   &   &   &   &    \\
D4 : & 0 & 1 & 2 & 3 &   &   & 6 &   &   &   & 10 \\
\hline
\end{tabular} \end{center}
\caption{N=2 Hanany-Witten setup}
\label{HananyWitten}
\end{table}
%
we parametrise 4 space by $u^1=x_4+ix_{10}$ and $u^2=x_5+ix_{6}$.
Note that D 4-brane positions $x^4$ and $x^5$ correspond to the $q^i$-variables,
whereas the NS 5-brane positions $x^{10}$ and $x^6$ are the conjugated $p^i$
variables.
Now we work out the two-cycle conditions on the two real
function $f^1=f(x^4,x^5,x^6,x^{10})$ and $f^2=g(x^4,x^5,x^6,x^{10})$. With
\begin{eqnarray}
M=\left(
\begin{array}{cc}
 f_{u^1} & f_{u^2}  \\
 g_{u^1} & g_{u^2}  \\
\end{array}\right)\nonumber
\end{eqnarray}
one can calculate the two-cycle equations which result in:
\begin{eqnarray}
    \{f,g\}\equiv 0\;\;\; \Rightarrow \;\;\;
    0&\equiv& g_4f_{10}-g_{10}f_4+g_5f_6-g_6f_5,\\  
    \det \, N\equiv 1\;\;\; \Rightarrow  \;\;\;
    0&\equiv& \; g_6f_4-g_{10}f_5-g_4f_6+g_5f_{10}.
\end{eqnarray}
In analysing these equations it is a simple task to verify that all functions 
$f$ and $g$ satisfying 
\begin{eqnarray}
  f_6 &=&g_{10}\;\;\;\;\;\;f_{10} = - g_6\nonumber\\
  f_4 &=&g_5   \;\;\;\;\;\;\;\;\, f_5 = - g_4\nonumber
\end{eqnarray}
do solve our equations. 
These are the ``Cauchy-Riemann'' differential equations which state that 
$f$ and $g$ are the real and imaginary part of a holomorphic function  
in the variables $v=x_4+ix_5$ and $s=x_6+ix_{10}$, respectively. 
Then we choose as our coordinates 
\begin{eqnarray}
    v &=& x_4+i\cdot x_{5}\nonumber \\
    t &=& e^{-s} = e^{-(x_6+i\cdot x_{10})}\nonumber 
\end{eqnarray}
to respect the compactness of the $x_{10}$ direction. 
Hence $f$ and $g$ fit into a holomorphic function in two variables 
$v$ and $t$.


Up to now we have shown, that there is a subclass of solutions to 
our equations, which coincides with the well known holomorphicity 
argument. But as they stand our equations are more general and we have 
to think about that difference. Nevertheless in a certain sense every 
geometrical two-cycle should be described by a holomorphic function, yet. 
That is to say in the whole variety of functions specifying the same 
geometrical two-cycle, there is a distinguished holomorphic function, i.e.
there is a lot of redundancy in the description, which could be exploited 
for constructing solutions, maybe. Therefore we are looking for a way of 
to mod out that redundancy.

This will be done by imposing some additional differential constraints 
in a generic way. To do that, recall the following properties of the 
matrix $N$:
\begin{enumerate}
\item By definition: $N=(-1)^d \bar{M}^{-1} M$.
\item As shown in appendix \ref{AppendixDerivationOfTheCycleEquation}, 
      N can be factorized through $a \in U(d)$
      \begin{eqnarray}
          N=\lambda^{-1} = \bar{a} a^{-1}.\nonumber
      \end{eqnarray}
\end{enumerate}
Note that this does not imply that $M$ must be unitary. But we want to 
choose M as close as possible to being unitary. 
We hope that this resolves the redundancy problem.

A unitary matrix $M$ does have orthonormalized rows and columns. Thus to 
begin with the construction of an unitary $M$ we want to orthogonalize the 
rows and columns of $M$. Since we know the expression for $MM^+$ in 
geometrical this is straightforward. Simply we have to require orthogonality 
of the gradients of $F$ and $G$ 
\begin{eqnarray}
       <\nabla\,F,\nabla\,G >\,=\,0.\nonumber
\end{eqnarray} 
Of course one has to be careful in doing that. One has to ensure that 
by requiring these additional properties, the common zero set is 
unchanged. In fact, this can be done without getting into trouble. 
If the length of these both gradients coincides, our equations reduce 
to the Cauchy-Riemann equations, indeed. 

There are problems in generalizing this nice looking observation to 
higher dimensional cycles but we hope that this construction 
works, too. 



\subsubsection{Supersymmetric 3-cycles}

Recall the characteristic $N=1$ brane configuration:\\[-25pt] 
\begin{center}
\parbox{10cm}
{
\refstepcounter{table}
\label{braneconf}
\font\bildfont=cmcsc10
\begin{center}
  \begin{tabular}{|c|ccccccccccc|}
  \hline
  NS : & 0 & 1 & 2 & 3 & 4 & 5 &   &   &   &   &    \\[-10pt]
  NS': & 0 & 1 & 2 & 3 &   &   & 6 & 7 &   &   &    \\[-10pt]
  D4 : & 0 & 1 & 2 &   & 4 &   & 6 &   &   &   & 10 \\
  \hline
  \end{tabular}
\end{center}
\center{{\bildfont\large Table} \thetable $:\;$ brane configuration.} 
}
\end{center}
%
Since the NS branes together with the D 4-branes build a $N=2$ subsystem,
two conjugated $(q,p)$ pairs are given by $(q_1,p_1)=(x^{10},x^{3})$ and 
$(q_2,p_2)=(x^6,x^5)$.
Then a single  NS brane as well as a single D 4-brane is 
automatically a supersymmetric
3-cycle, namely a supersymmetric 2-cycle in the $x^3$-$x^5$-$x^6$-$x^{10}$ space
times the  line $x^7={\rm const}$ in $x^4$-$x^7$ space.
The last pair of coordinates is fixed by the requirement
that also the NS' brane is a supersymmetric 3-cycle:
$(q_3,p_3)=(x^4,x^7)$. Note that with this choice the three coordinates 
$x^3$, $x^5$, $x^7$ of the D 4-branes are all momentum variables.
In summary, the complex structure of ${\mathbb{C}}^3$ takes the following form:
\begin{eqnarray}
    u^1 &=& x^{10}+i\cdot x^3\nonumber \\
    u^2 &=& x^4   +i\cdot x^7\nonumber \\
    u^3 &=& x^6   +i\cdot x^5\nonumber
\end{eqnarray}
Now we can work out the supersymmetric 3-cycle conditions on the
three functions\linebreak $f^1=f(x^3,x^4,x^5,x^6,x^7,x^{10})$,
$f^2=g(x^3,x^4,x^5,x^6,x^7,x^{10})$ and $f^3=h(x^3,x^4,x^5,x^6,x^7,x^{10})$.
First, the three Poisson brackets are given by the following set
of equations:
\begin{eqnarray}
   0\equiv\{ f,g\}&=&f_{10}g_3-f_3g_{10}+f_4g_7-f_7g_4+f_6g_5-f_5g_6,
     \nonumber\\
   0\equiv\{ f,h\}&=&f_{10}h_3-f_3h_{10}+f_4h_7-f_7h_4+f_6h_5-f_5h_6,
     \nonumber\\
   0\equiv\{ g,h\}&=&g_{10}h_3-g_3h_{10}+g_4h_7-g_7h_4+g_6h_5-g_5h_6.
   \label{poisson3}
\end{eqnarray}
The $\det \, N\equiv 1$ equation takes the following form:
\begin{eqnarray}
  0&\equiv&\left(f_4g_6-f_7g_5-f_6g_4+f_5g_7 \right)h_{10}
     +\left(g_{10}f_6-g_{3}f_5-f_{10}g_6+f_{3}g_5\right)h_4\nonumber\\
   &+&\left(f_{10}g_4-f_{3}g_7-g_{10}f_4+g_{3}f_7\right)h_6
     +\left(g_{3}f_4+g_{10}f_7-f_{10}g_7-f_{3}g_4\right)h_5\nonumber\\
   &+&\left(f_{10}g_5+f_{3}g_6-g_{3}f_6-g_{10}f_5\right)h_7
     +\left(f_6g_7-f_7g_6-f_4g_5+f_5g_4      \right)h_{3}.\label{detn3}
\end{eqnarray}
For a supersymmetric 3-cycles these four equations must be zero, but
not necessarily    identically, 
but in general only on the 3-cycle, i.e. modulo the ideal of
vanishing functions determined by $f$, $g$ and $h$.

One particular class of solutions for these equations is of course given
by all 3-cycles which are a supersymmetric 2-cycle in 
the $x^3$-$x^5$-$x^6$-$x^{10}$ space 
times the  line $x^7={\rm const}$ in $x^4$-$x^7$ space:
$\Sigma^{(3)}=\Sigma^{(2)}\times {\Bbb{R}}$.
The corresponding choice of functions is
$f=f(x^3,x^5,x^6,x^{10})$, $g=g(x^3,x^5,x^6,x^{10})$, $f$ and $g$ being real
and imaginary parts of a holomorphic function $F(x^3+ix^5,x^6+ix^{10})$,
and $h=x^7-{\rm const}$.

As a first and very simple check we can verify that flat, parallel 
M5-branes in their three possible asymptotic limits, namely being NS,  NS' or
D4-branes, are indeed  supersymmetric 3-cycles.
For example consider the $n$ parallel NS 5-brane, positioned at 
$x^6_i$, $x^7_i$ and $x^{10}_i$ ($i=1,\dots ,n$).
Hence the three functions $f$, $g$ and $h$ are given as
\begin{eqnarray}
f&=&\prod_{i=1}^n(x^6-x^6_i),\nonumber\\ 
g&=&\prod_{i=1}^n(x^7-x^7_i),\nonumber \\
h&=&\prod_{i=1}^n(x^{10}-x^{10}_i).
\end{eqnarray}
It is easy to show that all eqs.(\ref{poisson3}) and (\ref{detn3}) are
identically zero. The same is of course true for $n'$
parallel NS' 5-branes and
$k$ parallel D 4-branes.
In the following sections we will discuss more complicate brane intersections
and bent brane configurations.

\subsection{Supersymmetric 3-cycles for intersecting 
            branes and $N=1$ brane boxes}

\subsubsection{Branes as quaternionic coordinates}

In the following sections we like to construct the defining equations
$f$, $g$ and $h$ for  those supersymmetric 3-cycles, which correspond to
intersecting  NS, NS' and D 4-branes, and in particular for those,
which correspond to $N=1$ brane box configurations.
For this purpose we would like to introduce three types of `coordinates',
called $s$, $s'$ and $v$, which denote the asymptotic positions 
in ${\Bbb{C}}^3$ of
the NS, NS' and D 4-branes respectively.
These `coordinates' should be one the same footing as the complex variables
$s=x^6+ix^{10}$ and $v=x^4+ix^5$ of the $N=2$ (NS-D4) brane configurations.

To achieve this aim we will now extend the dimension of the space
by including also the directions $x^2$ and $x^8$. This means that we are
now dealing with supersymmetric 4-cycles which are embedded into the
space ${\mathbb{C}}^4$, which is spanned by the directions (2,3,4,5,6,7,8,10).
All our branes now fill 4 dimensions of this eight dimensional space:
their world volumes completely fill $x^2$, and they are all positioned
at $x^8=0$. That means that the 4-cycles, which correspond to the
brane boxes of the NS, NS' and D 4-branes are in fact nothing else
than  supersymmetric 3-cycles times the line $x^8=0$.
As discussed in detail above,
we could add yet another type of NS-branes, called NS'' branes, with
world volumes along the (3,4,6,8)-directions and positions in
the (2,5,7,10) space. 
Considering intersections of all four types of branes (NS-NS'-NS''-D4)
one can construct brane cube models, where the D 4-branes are now finite
in the directions $x^2$, $x^4$ and $x^6$.
These brane cubes provide two-dimensional gauge theories with (1,1)
supersymmetry.
A generic brane cube configuration corresponds to a supersymmetric 4-cycle,
which is not a direct product of supersymmetric 3-cycle times ${\Bbb R}$.

The positions of the branes in ${\Bbb{C}}^4$ can now nicely described
by introducing quaternionic numbers. A general quaternion $q\in {\Bbb{H}}$
has the
structure
\begin{equation}
      q=q^0\sigma_0+q^1\sigma_1+q^2\sigma_2+q^3\sigma^3,
\end{equation}
where $\sigma_0={\unity}_2$ and the $\sigma_i$ ($i=1,2,3$) are
the Pauli matrices, satisfying $\sigma_i\sigma_j=\epsilon_{ijk}\sigma_k$.
Clearly, a quaternion is zero, $q=0$, if all its components
$q_i$ ($i=0,\dots ,3$) are vanishing. Alternatively, we can also define
the quaternions via two complex numbers $z_1=q^0+iq^1$ and $z_2=q^2-iq^3$ as
$q=z_1+jz_2$,
where $i=\sigma_1$, $j=\sigma_2$ and $k=i\cdot j=\sigma_3$.

Now we can associate to every brane a particular quaternion $q$, which describes
its asymptotic position in ${\Bbb{C}}^4$, and hence is a function of the
position variables of every brane:
\begin{eqnarray}
NS\,:\; \quad q_{NS}&=&q(x^6,x^7,x^8,x^{10}),\nonumber\\
NS':  \quad q_{NS'}&=&q(x^4,x^5,x^8,x^{10}),\nonumber\\
D4\;\,:\, \quad q_{D4}\,\,&=&q(x^3,x^5,x^7,x^{8}).
\end{eqnarray}
The four defining equations $f^m(x^2,x^3,x^4,x^5,x^6,x^7,x^8,x^{10})$ 
($m=1,\dots ,4$) for the 4-cycle can be now simply written in terms
of a single quaternionic function  function $F(q_{NS},q_{NS'},q_{D4})$:
\begin{equation}
    F(q_{NS},q_{NS'},q_{D4}) = f^1(x^i)+if^2(x^i)+jf^3(x^i)+kf^4(x^i).
\end{equation}
Of course, for a general function $F(q_{NS},q_{NS'},q_{D4})$ 
one still has to verify
whether the 4-cycle is supersymmetric. This is not automatic unlike
the case of the supersymmetric 2-cycles, where every holomorphic function
corresponds to a supersymmetric 2-cycle.
Specifically, as
discussed in section \ref{SectiondCycleEquation},
the supersymmetry conditions are given by the requirement that
six Poisson 
brackets $\{ f^m,f^n\}$ plus $(\det N-1)$ have to vanish (modulo the
ideal of vanishing functions determined by the zero locus
of the $f^m$).
In addition, since we want the supersymmetric 4-cycle $\Sigma^{(4)}$ 
to be of the
form $\Sigma^{(4)}=\Sigma^{(3)}\times{\mathbb{R}}_{x^8=0}$, the function
$F(s,s',v)$ has to be chosen in such a way that the 
common zero locus of the $f^m$ always contains the line
$x^8=0$.
In principle it is also possible to obtain the three 3-cycle equations
$f$, $g$ and $h$ by solving one of the four equations $f^m$ with respect
to $x^8$ and substituting the result into the remaining equations.

To understand this procedure of constructing supersymmetric 3-cycles
let us first consider case of classical brane configurations which are
not bent by quantum effects.
To describe flat branes we introduce the following quaternionic coordinates
in analogy to the complex variables $s$ and $v$\footnote{The NS''-brane
corresponds to the quaternion $s''=x^{10}+ix^5+jx^7-kx^2$.}:
\begin{eqnarray}
NS\,:  \quad  s\,&=&x^6+ix^{10}+jx^7-kx^8,\nonumber\\
NS': \quad s'&=&x^{4}+ix^5+jx^{10}-kx^8,\nonumber\\
D4\;\;:  \quad  v\,&=&x^3+ix^5+jx^7-kx^8.
\end{eqnarray}
A single NS brane is a supersymmetric
4-cycle simply defined by the equation
$F=s=0$ and likewise for the other branes.
Next consider the triple intersection of $n$ parallel
NS branes with $n'$ parallel NS' branes and $k$ parallel D 4-branes. 
This configuration
corresponds to un-bent NS and NS' branes. In the language of
field theory it leads to finite $N=1$ gauge theories.
The associated quaternionic function $F$ is given by the following
polynomial:
\begin{equation}
F(s,s',v)=\prod_{i=1}^n(s-s_i)\prod_{j=1}^{n'}(s'-s_j')\prod_{l=1}^k(v-v_l).
\end{equation}
Here $s_i$, $s_j'$ and $v_l$ are constant quaternionic numbers with
zero $\sigma_3$-component, which denote the positions
of the three types of branes.
It is not a difficult but a tedious calculation to show that this function
$F$ corresponds to a supersymmetric 4-cycle. However note that the 
supersymmetric 4-cycle equations are
not identically satisfied but only on the branes themselves, i.e.
only modulo the ideal of vanishing functions of the $f^m$.

\subsubsection{Uniform Bending -- Sewing of $N=2$ models}
\label{sectionUniformBending}

Now we will construct the supersymmetric 3-cycles which correspond to those
$N=1$ brane boxes which can be obtained via the sewing or superposition
of two $N=2$ subsystems. As explained in section (3.1), this means
that all the NS branes as well as all the NS' branes are bent in an
uniform way.
 
In general, 
the bending of the NS and NS' branes should be parametrized by the $x^3$
position of the D 4-branes, where $x^3$ is nothing else that the parameter
which is associated to the Coulomb branch in three dimensions.
In addition, we roughly expect that the 
bending of the NS brane is encoded in the functions $x^6(x^3)$ and
$x^{10}(x^3)$, and analogously, the bending of the NS' branes
is determined by  $x^4(x^3)$ and $x^{10}(x^3)$. Since $x^3$ takes 
in four dimensions the role
of $\Lambda_{QCD}$, $x^4$, $x^6$ and $\cos{x^{10}}$ 
($x^{10}$ is periodic!) should be logarithmic functions of $x^3$.

For the case of uniform bending we can be  much more explicit.
Consider first the uniform bending of the NS brane caused by
$k$ D 4-branes. From the $N=2$ models we know that the perturbative
bending is described by a two dimensional 
Laplace equation  with the holomorphic, logarithmic solution
$x^6+ix^{10}=k\log (x^3+ix^5)$.
In the same way, for the other $N=2$ subsystem, NS'-- k'D4, the 
following perturbative solution for the
bending holds: $x^4+ix^{10}=k'\log ( x^7+ix^3)$.
This behaviour now suggest that we define the following quaternionic
coordinates which describe the asymptotic  positions of the bent branes
in a correct way:
\begin{eqnarray}
NS:  &{}&\quad t\;=e^{x^6}\cos x^{10}+ie^{x^6}\sin x^{10},
     \nonumber\\
NS': &{}&\quad t'=e^{x^4}\cos x^{10}+je^{x^4}\sin x^{10},
     \nonumber\\
D4:  &{}&\quad v\,=x^3+ix^5+jx^7-kx^8.\label{ttv}
\end{eqnarray} 
Sewing together
the perturbative bending of the 
two $N=2$ subsystems provides us with the following quaternionic
function for the supersymmetric 3-cycle, which corresponds
to the  simple brane box shown in figure \ref{figureN2BraneBox}:
\begin{figure}[htb]
  \makebox[16cm]{
    \epsfxsize=6cm
    \epsfysize=6cm
    \epsfbox{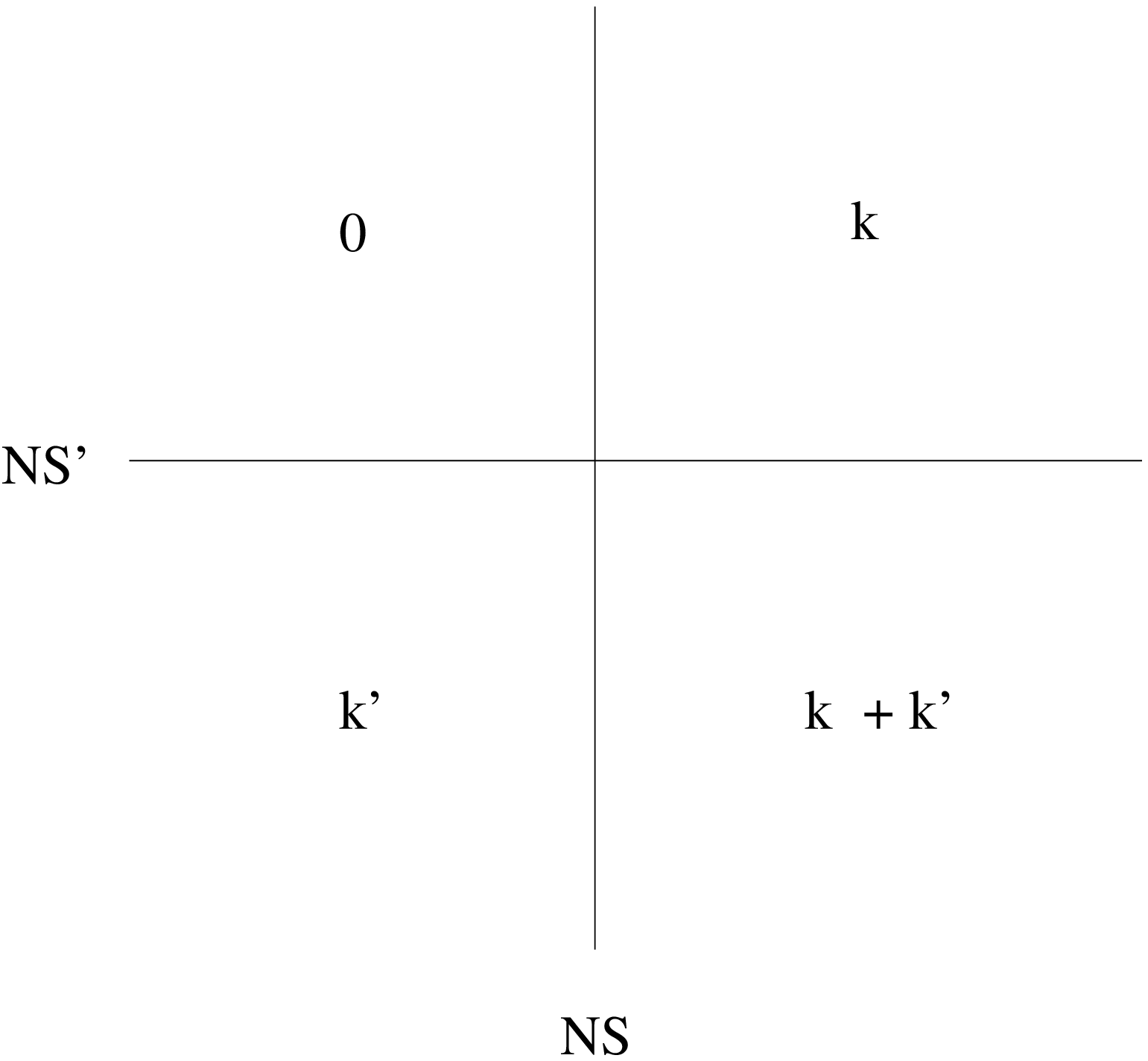}
  }
  \caption{A simple brane box with uniform bending}
  \label{figureN2BraneBox}
\end{figure}                
%
For $k=k'=1$ the quaternionic function simply takes the form
\begin{equation}
F(t,t',v)=\lbrack t-v\rbrack\lbrack t'-v\rbrack
=0.\label{n=1bend}
\end{equation}
Similar one can write down an expression for arbitrary $k$ and $k'$.
It is possible to show that this function satisfies
the conditions for a supersymmetric cycle.
The vanishing locus which is defined by $F(t,t',v)$ is a true 3-cycle;
it consists out of two branches, namely the superposition
of the curve $t-v^k=0$, which is a 2-cycle in the $3-5-6-10$-directions
times the $x^4$-axis,  
together with the curve $t'-v^{k'}=0$, which represents a 2-cycle, now
in the directions $3-4-7-10$ times the $x^6$-axis. 

After having understood the most simple  $N=1$ brane box 
with uniform bending (see figure \ref{figureN2BraneBox})
we can now construct the non-perturbative,
supersymmetric 3-cycle equations which describe
the generic $N=1$ brane box with
 uniform bending situation. It is given by the superposition
of two $N=2$ subsystems: the first one consists out of $n$ NS 5-branes with
$k_\alpha$ D 4-branes suspended between the NS branes 
(see figure \ref{figureBraneConfiguration}).
The second $N=2$ subsystem has the same structure, but now $n'$ NS' branes with
$k_{\alpha '}'$ suspended D 4-branes. After sewing 
together these two subsystems,
the $N=1$ brane box has the form shown in figure 
\ref{figureUniformBraneBoxWeb}.
Now recall that, non-perturbatively, every $N=2$ system of this kind
is characterized by the complex 2-cycle polynomial eq.(\ref{n=2pol}).
Then the sewing procedure simply corresponds to the multiplication of
the two $N=2$ polynomials, where we replace the complex variables by the
corresponding quaternionic variables. In this way
 we get a supersymmetric 3-cycle
which consists out of two branches, namely the direct sum 
\begin{equation}
\Sigma^{(3)}_{n,n',k_\alpha,k_{\alpha '}'}= (\Sigma^{(2)}_{n,k_\alpha}
\times{\mathbb{R}})\oplus (\Sigma^{(2)}_{n',k_{\alpha '}'}\times{\mathbb{R}})
.
\end{equation}
Note that the two superposed 3-cycles have a common volume in the
3-10 space.
In general the quaternionic 3-cycle equations will have the following 
structure:
\begin{equation}
\Sigma^{(3)}:\quad
F(t,t',v)=\lbrack p_{k_0}(v)t^n+\dots 
+p_{k_n}(v)\rbrack\lbrack 
p_{k_0'}(v){t'}^n+\dots 
+p_{k_{n'}'}(v)\rbrack .\label{n=3pol}
\end{equation}
This expression can be expanded and one obtains a polynomial of the
following structure:
\begin{equation}
F(t,t',v)=\sum_{\alpha=0}^n\sum_{\alpha'=0}^{n'}p_{k_\alpha}(v)
p_{k_{\alpha '}'}(v)
\, t^{n-\alpha}\, {t'}^{n'-\alpha '}.\label{unifpol}
\end{equation}
Note that the degree of the polynomial in
$v$ in front of each term $t^{n-\alpha}\, {t'}^{n'-\alpha '}$
 precisely agrees with the number of
D 4-branes in each box $\lbrack \alpha,\alpha '\rbrack$.

\begin{figure}[htb]
  \makebox[16cm]{
    \epsfxsize=6cm
    \epsfysize=6cm
    \epsfbox{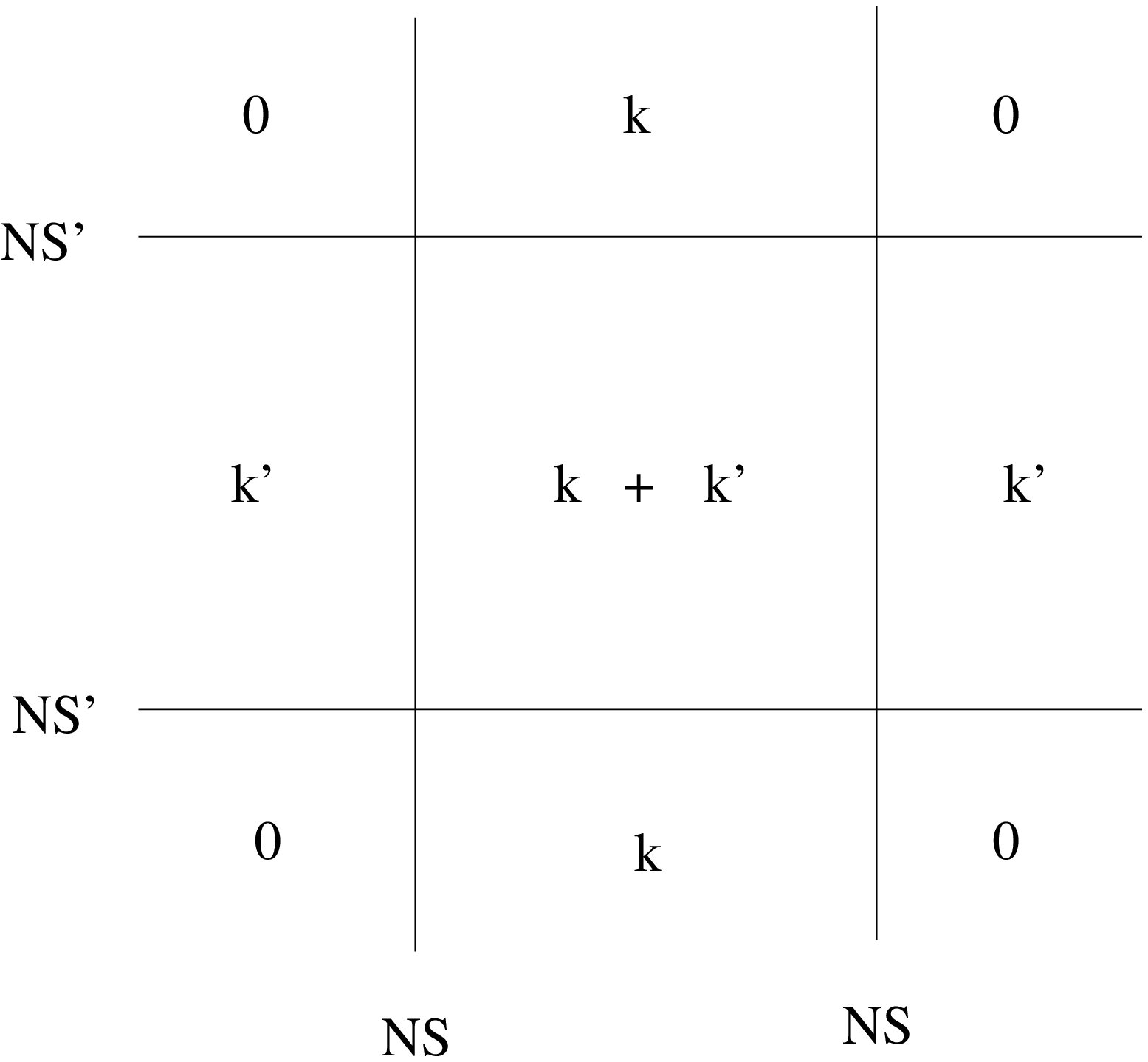}
  }
  \caption{Sewing of two pure $N=2$ Yang-Mills theories}
  \label{figureUniformBending}
\end{figure}

For example the sewing of two pure $N=2$ gauge theories with $G=SU(k)$
and $G'=SU(k')$ leads to 
a $N=1$ gauge theory with $N_c=N_f=k+k'$ (see figure 
\ref{figureUniformBending}).
%
The corresponding 3-cycle equations are then simply given in terms
of the product of two Seiberg-Witten elliptic curves of genus $(k-1)$
resp. $(k'-1)$. This strongly suggests that the instanton numbers
of the pure $N=2$ Yang-Mills theory with gauge group $SU(k)$ are intimately
related to those of SUSY QCD with $G=SU(2k)$ and $N_f=2k$.

At the end of this section let us compute the perturbative
running of the $N=1$ gauge coupling constant.
A priori we deal with two different Coulomb branches parametrized
by $x^3+ix^5$ resp. by $x^3+ix^7$. In the following we will
consider the common direction, $x^3$, and freeze the other directions,
i.e. $x^5=x^7=0$. Now
consider the box  $\lbrack \alpha, \alpha '\rbrack$ with the corresponding
gauge group $SU(k_\alpha+k_{\alpha '}')$.
From eq.(\ref{n=1bend}) we derive that
\begin{eqnarray}
x^4_{\alpha '+1}-x^4_{\alpha '}&=& L+(k_{\alpha '+1}'+k_{\alpha '-1}'-
2k_{\alpha '}')\log x^3,\nonumber\\
x^6_{\alpha +1}-x^6_{\alpha }&=& L+({k}_{\alpha +1}+{k}_{\alpha -1}-
2{k}_{\alpha })\log x^3,
\end{eqnarray}
where $L$ is the classical distance between the NS and NS' branes.
Then using eq.(\ref{n=1gaugec}), the gauge coupling constants exhibits
the following running behaviour:
\begin{eqnarray}
\frac{1}{g_{\alpha,\alpha'}^2}&=&
({g_s})^{-1}\biggl(L^2+L( k_{\alpha '+1}'+k_{\alpha '-1}'+     
{k}_{\alpha +1}+{k}_{\alpha -1}-2k_{\alpha '}'-2{k}_{\alpha })\log x^3
\nonumber\\
&+&(k_{\alpha ' +1}'+k_{\alpha '-1}'-
2k_{\alpha '}')({k}_{\alpha +1}+{k}_{\alpha -1}-
2{k}_{\alpha })(\log x^3)^2
\biggr).
\label{n=1gaugerun}
\end{eqnarray}
Since $N_c=k_\alpha+k_{\alpha '}'$ and 
$N_f=k_{\alpha '+1}'+k_{\alpha '-1}'+
k_{\alpha '}'+{k}_{\alpha +1}+{k}_{\alpha -1}+
{k}_{\alpha }$, the coefficient in front of $\log x^3$ precisely
agrees with the one-loop $N=1$ $\beta$-function coefficient
$b_{N=1}=-3N_c+N_f$\footnote{It was already observed in refs.\cite{haur,AB}
that the brane box models with uniform bending lead to the correct
$N=1$ $\beta$-function coefficients.}.

\subsubsection{General $N=1$ brane boxes}

In the last section we have discussed already a quite large class of
$N=1$ gauge theories, namely those $N=1$ models with $N_f\geq N_c$ 
can be obtained by sewing $N=2$ brane configurations.
On the other hand, $N=1$ gauge theories with $N_f<N_c$ like pure $N=1$
Yang-Mills
and also models with chiral fermions are more general
and cannot be obtained by the $N=2$ sewing procedure.
Of course these models are very interesting to study dynamical supersymmetry
breaking and the effect of anomalies.
In general, we expect that a brane box which corresponds in field theory
to a model without vacuum at finite value of the
moduli, like supersymmetric QCD
with $0<N_f<N_c$, leads to a 3-cycle which does not
satisfy the minimal area requirement.
In this case the 3 Poisson brackets may still be zero, i.e. $N\equiv N^T$, but  
$Re(\det \, N)$ is non-vanishing.
Similarly, in chiral $N=1$ gauge theories with
dynamical supersymmetry breaking we expect a stable non-supersymmetric
ground state. That is the $det$ requirement will be satisfied, while
the cyle won't be a Lagrangian submanifold anymore. 
If furthermore the model is anomalous, the 3-cycle should not exist
at all.

In the following we like to propose a specific ansatz for the 3-cycle equations
for a general $N=1$ brane box model. We will again use the quaternionic
formalism with quaternions $t$, $t'$ and $v$ (see eq.(\ref{ttv})).
Motivated by the previous
discussions, our ansatz will consist out of an polynomial
in these variables, where the degree of the polynomial in $t$ ($t'$)
corresponds to the number of NS (NS') branes in the
corresponding brane box. 
Hence, for
a general brane box as shown in figure (\ref{figureGeneralBraneBoxWeb})
the quaternionic 3-cycle equations are assumed to take the following
form
\begin{equation}
\Sigma^{(3)}:\quad
F(t,t',v)=\sum_{\alpha=0}^n\sum_{\alpha'=0}^{n'}p_{k_{\alpha,\alpha '}'}(v)
\, t^{n-\alpha}\, {t'}^{n'-\alpha '}.\label{n=3pol1}
\end{equation}
$p_{k_{\alpha,\alpha '}}(v)$ is a polynomial in $v$ whose degree
is given by the number $k_{\alpha,\alpha '}$ of D 4-branes in each box
$\lbrack \alpha,\alpha '\rbrack$.
As  already said, for a brane box which corresponds in field theory
to an anomaly free gauge theory with supersymmetric vacuum, one should be
able to proof \cite{workinpro} that this
polynomial  provides a supersymmetric 3-cycle.
At the moment it is however not possible for us to show this in general; 
the main technical difficulty
is the observation that the supersymmetric 3-cycle equations must be satisfied
only modulo the ideal $I({\Bbb{V}})$ of functions vanishing on the 3-cycle
${\Bbb{V}}(f,g,h)$.
Note however that in 
case of uniform bending the polynomial eq.(\ref{n=3pol1}) takes the
form of eq.(\ref{unifpol}), namely it factorizes as in eq.(\ref{n=3pol}),
and the supersymmetric 3-cycle equations are  satisfied.

A particularly interesting case is pure supersymmetric
QCD with gauge group $SU(k)$. Here the 3-cycle polynomial 
should have the following structure.
\begin{equation}
F(t,t',v)=
\sum_{\alpha=0}^2\sum_{\alpha'=0}^{2}
\, t^{2-\alpha}\, {t'}^{2-\alpha '}+p_k(v) \, t \, t',
\end{equation}
where $p_k(v)$ is a polynomial in $v$ of degree $k$.
For finite $R_3^{IIB}$ which includes the decompactification limit to
four dimensions, $R_3^{IIB}\rightarrow\infty$, there exist a supersymmetric
vacuum in field theory such the supersymmetric 3-cycle equations
should be satisfied for this ansatz.
On the other hand, in the 3-dimensional limit, $R_3^{IIB}\rightarrow 0$,
the supersymmetric 3-cycle equations should be violated, since
there is no supersymmetric vacuum in 3-dimensional pure Yang-Mills gauge
theory \cite{workinpro}.

\section{Conclusions}

We have shown that SUSY 3-cycles play a
similar role in $N=1$ SUSY gauge theories as the Seiberg-Witten 
curve in $N=2$ in
the sense that their geometry encodes the holomorphic information
about the gauge theory.
Especially we expect the superpotential to correspond to the volume
and the couplings on the Coulomb branch (if present) to the
periods of the cycle \cite{workinpro}. We were able to construct these cycles
for gauge theories that satisfy the uniform bending requirement
of \cite{gimongremm}. The tools we used in establishing these
cycles should be useful for the more general cases as well.\\[0.2cm]

\underline{Note added:} See the footnote on page \pageref{Tragik}.\\[0.2cm]

\noindent{\bf Acknowledgements:}

\noindent 
 Work partially supported by the E.C. project
ERBFMRXCT960090 and the Deutsche Forschungs Gemeinschaft.
We like to thank Douglas Smith for collaboration during early stages
of this work. In addition we acknowledge useful discussions with
A. Hanany, A. Krause, Y. Oz, R. Reinbacher, S. Theisen and A. Zaffaroni.

%
%

\begin{appendix}

\setcounter{section}{0}


\refstepcounter{section}
\section*{Appendix \thesection : Derivation of the d-cycle equations}
\label{AppendixDerivationOfTheCycleEquation}
\addtocontents{toc}{{\bf Appendix \thesection:} Derivation of the 
                      d-cycle equations\hfill {\bf\thepage}\\[0.25cm]}

Consider the
embedding map $i:d-{\rm cycle}\longrightarrow M^{2d}_{\mathbb{R}}$ and 
the two conditions:
\begin{eqnarray}
      i^\ast \MyIm \Omega &=& 0  \;\;\;\;  {\rm volume\;\; minimizing}
      \nonumber\\
      i^\ast \omega &=& 0      \;\;\;\;  {\rm Lagrangian\;\; submanifold}
      \nonumber 
\end{eqnarray}
 With $\Omega=du^1\wedge\ldots\wedge du^d$ the requirement of 
minimal volume reads
\begin{eqnarray}
0 = i^\ast\MyIm\Omega&=&\MyIm (du^1(\xi_1,{\ldots},\xi_d)\wedge 
                               \ldots\wedge 
                               du^d(\xi_1,{\ldots},\xi_d))\nonumber\\ 
  &=&\MyIm( \epsilon_{i_1{\ldots}i_d}u^{i_1}_{\xi_1}u^{i_2}_{\xi_2}{\ldots}u^{i_d}_{\xi_d})\; 
             d\xi^1\wedge{\ldots}\wedge d\xi^d\nonumber\\
\Rightarrow\;\;\;\;
0
&=&\frac{1}{2i}(\epsilon_{i_1{\ldots}i_d}u^{i_1}_{\xi_1}{\ldots}u^{i_d}_{\xi_d} -
                \epsilon_{i_1{\ldots}i_d}\bar{u}^{i_1}_{\xi_1}{\ldots}
                               \bar{u}^{i_d}_{\xi_d} )\nonumber\\
  &=&\frac{1}{2i}(\epsilon_{i_1{\ldots}i_d}u^{i_1}_{\xi_1}{\ldots}u^{i_d}_{\xi_d}
                 -\epsilon_{i_1{\ldots}i_d} N^{i_1}_{j_1}{\ldots}N^{i_d}_{j_d} u^{j_1}_{\xi_1}
                             {\ldots}u^{j_d}_{\xi_d})\nonumber\\
  &=&\frac{1}{2i}(\epsilon_{i_1{\ldots}i_d}
                 -\epsilon_{j_1{\ldots}j_d}N^{j_1}_{i_1}{\ldots}N^{j_d}_{i_d}) 
                  u^{i_1}_{\xi_1}{\ldots}u^{i_d}_{\xi_d}\nonumber\\
  &=&\frac{1}{2i}\left(1-\det\,N\right)
      \cdot\frac{\partial (u^1,{\ldots},u^d)}{\partial (\xi_1,{\ldots},\xi_d)}
     \nonumber
\end{eqnarray}
which  yields
\begin{eqnarray}
    \det \, N|_{{\Bbb{V}}(f^1,{\ldots},f^n)} = 1\;\;\;\; 
               &{\rm or\; for\; short}&\;\;\;\;
             \det\, N \equiv 1.\nonumber
\end{eqnarray}
For the calculation of the det-equation the following relation is useful.
\begin{eqnarray}
    \det\, N &\equiv& 1\;\;\;\Leftrightarrow\;\;\; 
    \det\, M - (-1)^d \det\, \bar{M}\equiv 0\nonumber
\end{eqnarray}

Now we turn to the second equation. 
Choosing the canonical K\"ahler (symplectic)
form $\omega = \frac{1}{2i}\sum\limits_{i} du^i\wedge d\bar{u}^i$, 
the pull back operation results in 
\begin{eqnarray}
  0 = i^\ast\omega &=&\frac{1}{2i}\sum\limits_{i}du^i\left(\xi_1\ldots\xi_d\right)
        \wedge d\bar{u}^i\left(\xi_1\ldots\xi_d\right)\nonumber\\
    &=& \frac{1}{2i}\sum\limits_{i}\left(\sum\limits_k u^i_{\xi_k}d\xi_k\right)
        \wedge\left(\sum\limits_l\bar{u}^i_{\xi_l}d\xi_l\right)\nonumber\\
    &=& \frac{1}{2i}\sum\limits_{k<l}\sum\limits_{i} \left[
        u^i_{\xi_k}\bar{u}^i_{\xi_l} -
        u^i_{\xi_l}\bar{u}^i_{\xi_k}   
        \right] d\xi_k\wedge d\xi_l\nonumber
\end{eqnarray}
\begin{eqnarray}
\Rightarrow\;\;\; 0 &=& \sum\limits_{i} \left[
        u^i_{\xi_k}\bar{u}^i_{\xi_l}-
        u^i_{\xi_l}\bar{u}^i_{\xi_k}   
        \right] \nonumber\\
&=& \sum\limits_{i} \left[
        u^i_{\xi_k}\left(\sum\limits_m N^i_m u^m_{\xi_l}\right)-
        u^i_{\xi_l}\left(\sum\limits_m N^i_m u^m_{\xi_k}\right)   
        \right] \nonumber\\
&=& \sum\limits_{i}
        u^i_{\xi_k}\left(\sum\limits_m N^i_m u^m_{\xi_l}\right)-
    \sum\limits_{i}
        u^i_{\xi_l}\left(\sum\limits_m N^i_m u^m_{\xi_k}\right) 
      \nonumber\\
&=& \sum\limits_{i}
        u^i_{\xi_k}\left(\sum\limits_m N^i_m u^m_{\xi_l}\right)-
    \sum\limits_{m}\left(\sum\limits_i
        u^i_{\xi_l} N^i_m\right) u^m_{\xi_k} 
      \nonumber\\
&=& \sum\limits_{i}
        u^i_{\xi_k}\left(\sum\limits_m N^i_m u^m_{\xi_l}\right)-
    \sum\limits_{i}\left(\sum\limits_m
        u^m_{\xi_l} {N^T}^i_m\right) u^i_{\xi_k} 
      \nonumber\\
&=& \sum\limits_{i,m}\left(N^i_m-{N^T}^i_m\right)u^i_{\xi_k} u^m_{\xi_l}
      \nonumber
\end{eqnarray}
which is  satisfied if we set $N\equiv N^T$.  
However, as it stands, this requirement is sufficient, only. Now we intend to 
give a proof that the condition is necessary, too.

To proof  $N\equiv N^T$ we remember some facts from
symplectic geometry especially various ways of characterising Lagrangian
planes in symplectic vector spaces. The utility of this investigation rest
on the simple observation that our conditions on the d-cycle to be a 
special Lagrangian submanifolds are in fact conditions on its tangent 
bundle, i.e. Lagrangian planes locally.\\[0.1cm]

To begin with, we consider a complex vector space ${\mathbb{C}}^d$ furnished 
with a Hermitian structure
\begin{eqnarray}
   <x,y>\, = \sum\limits_i x_i\bar{y}_i = g(x,y) + i\,\sigma(x,y) \nonumber
\end{eqnarray}
which splits into an Euclidean metric $g$ and a symplectic form $\sigma$. 
One can check that $\sigma$ coincides with
\begin{eqnarray}
  \omega = \frac{1}{2i}\sum\limits_i du^i\wedge d\bar{u}^i. \nonumber
\end{eqnarray}
given before. Therefore we identify both 
objects. 
The two-form $\omega$ is non degenerated, antisymmetric and bilinear. 
With help of $\omega$ we can define the notion of symplectic orthogonality.
\begin{defn}
  The orthogonal complement of a vector subspace $E\in {\mathbb{C}}^d$ is
  defined by
  \begin{eqnarray}
         E^\perp  = \{ x\in {\mathbb{C}}^d \mid \omega(x,E) = 0 \} \nonumber
  \end{eqnarray}
\end{defn} 
In the special case that $E=E^\perp$ we call $E$ a Lagrangian plane. 
Obviously on a Lagrangian plane the symplectic form restricts to zero. So we
recognize the content of the constraint $i^\ast\omega=0$. It simply states 
that all tangent spaces to the supersymmetric cycle are Lagrangian planes 
embedded in the tangent space of the embedding space. Here we collect some 
facts:
\begin{enumerate}
\item $Sp(E)$ operates transitively on Lagrangian planes
\item Since $U(d)$ preserves the Hermitian form, it is contained in $Sp(E)$.
\item By $\Lambda(d)$ we denote the Gra{\ss}mannian of Lagrangian planes 
\item $\lambda\in\Lambda(d)$ is characterized by choosing an orthonormal 
       basis $(a_1,\ldots ,a_n)$ with respect to the Euclidean metric $g$.
      But then it is orthonormal with respect to the Hermitian form, too: 
      \begin{eqnarray}
          < a_i,a_j >\, = g(a_i,a_j) + i\,\omega(a_i,a_j) 
          \buildrel !\over = \delta_{ij}, \nonumber
      \end{eqnarray}
      i.e. the matrix $a=(a_1,\ldots , a_n)$ is unitary. The other direction 
      works, too. 
      Hence 
      \begin{eqnarray}
        \lambda\in \Lambda(d) \;\; \Leftrightarrow\;\; \exists\, a\in U(d),\; 
        \lambda = a({\mathbb{R}}^d) \nonumber
      \end{eqnarray}
      \begin{figure}[htb]
      \makebox[16cm]{
         \epsfxsize=6cm
         \epsfysize=6cm
         \epsfbox{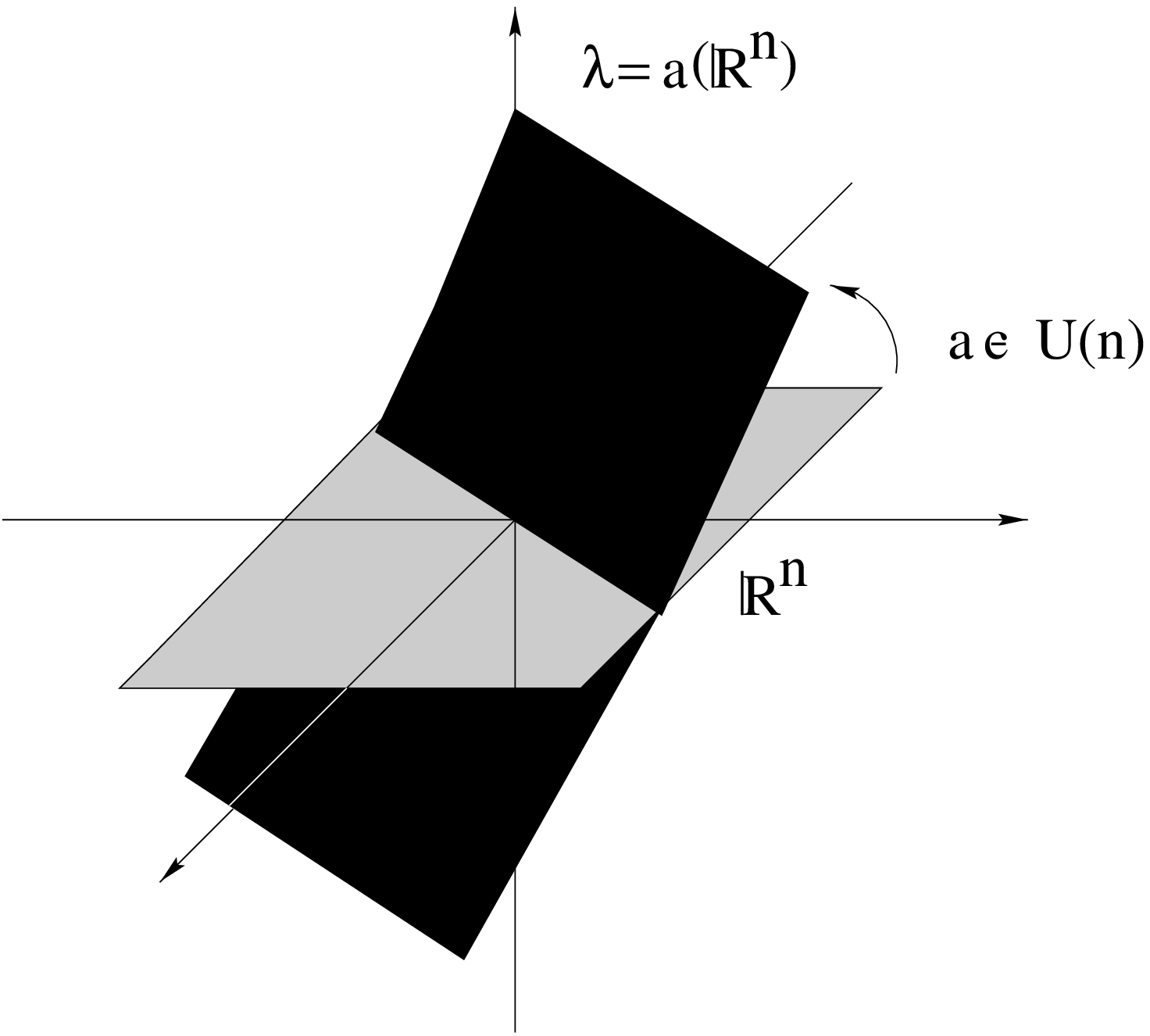}
      }
      \caption{The operation of $a$ on Lagrangian planes}
      \label{LagrangianPlanes}
      \end{figure}
      %
\item Obviously each Lagrangian plane will be stabilized by any element 
      in $O(n)$, i.e. we can regard the Gra{\ss}mannian of Lagrangian planes 
      as the quotient space
      \begin{eqnarray}
         \Lambda(d) = \frac{U(d)}{O(d)} \nonumber
      \end{eqnarray} 
\end{enumerate}
How can we define a projection from $U(d)$ onto $\Lambda(d)$? We observe 
that two elements $a$ and $a'$ determine the same Lagrangian plane, iff 
\begin{eqnarray}
 \lambda = a({\mathbb{R}}^d) ={a'}({\mathbb{R}}^d) 
             \Leftrightarrow\;\;\;
       a\bar{a}^{-1} = {a'}{\bar{a}}^{\prime -1}, \nonumber
\end{eqnarray}
which is constant on the $O(d)$-orbits of the fibration.
Now we can identify $\Lambda(d)$ with the image of the projection map
\begin{eqnarray}
    \pi : U(d) &\rightarrow&  \Lambda(d)\nonumber\\
           a     &\mapsto&    \lambda = a\bar{a}^{-1} \nonumber
\end{eqnarray}
By abuse of language we denote the matrix representative $a\bar{a}^{-1}$ of 
the Lagrangian plane $\lambda=a({\Bbb{R}}^n)$ by $\lambda$ again. 
But how can we associate the geometrical object with this artificial matrix
representative? The  connection between the matrix $\lambda$ on the one side and the concrete Lagrangian plane $\lambda$ on the other side is given through the central 
equation
\begin{eqnarray}
   x\in\lambda\;   \Leftrightarrow\;  x=\lambda\bar{x} \nonumber  
\end{eqnarray}
In the last formula we recognize the familiar equation  
(\ref{lemma}).
But now we know, that we can represent $\lambda$ as $\lambda=a\bar{a}^{-1}$
and this yields straight forward
\begin{eqnarray}
   \lambda^{+} &=& {\bar{a}^{-1^{+}}}a^{+} = {a^{-1}}^T a^{-1} = \bar{a}a^{-1}
                = \bar{\lambda} \nonumber\\
\Rightarrow\;\; \lambda^T &=& \lambda \nonumber
\end{eqnarray}
But then we can finally conclude by identifying  $\lambda = N^{-1}$ and 
performing some mild manipulations that
\begin{eqnarray}
      N \equiv N^T\nonumber
\end{eqnarray}
\rightline{\qed}

\pagebreak
\refstepcounter{section}
\section*{Appendix \thesection: Some facts from Hamiltonian dynamics}
\label{AppendixLiouville}
\addtocontents{toc}{{\bf Appendix \thesection:} Some facts from 
                    Hamiltonian dynamics\hfill 
                    {\bf\thepage}\\[0.25cm]}

\begin{thm}{(Liouville)}
{  Suppose $(f^1,\ldots, f^d)$ is a set of smooth functions on a symplectic 
   manifold $M^{2d}$ that are pairwise in involution, i.e. $\{f^i,f^j\}=0$. 
   \linebreak Let $M_{\xi}$ be the joint level surface determined by a 
   system of 
   equations\linebreak  $f^1(x)=\xi_1,\ldots,f^d(x)=\xi_d$. Suppose the 
   functions are 
   functionally independent on $M_\xi$ (that is, the gradients of the 
   functions are linearly independent at each point of $M_\xi$). Then the 
   following assertions are true:
   \begin{enumerate}
     \item The level surface $M_{\xi}$ is a smooth $n$-dimensional 
           submanifold that is invariant with respect to the flows 
           determined by the vector fields $X_{f^i}$.
     \item The connected components of $M_\xi$ are diffeomorphic to 
           $T^k\times{\mathbb{R}}^{d-k}$.
     \item If $M_{\xi}$ is compact and connected, then it is diffeomorphic 
           to the $d$-dimensional torus $T^d$
   \end{enumerate}
}
\end{thm}
Now we want to show how we utilize this theorem for our purposes. At first 
we identify the functions $(f^1,\ldots, f^d)$ as the defining equations of 
our searched for intersection. Then we start with such functions $f_i$, 
such that the gradients are linear independent everywhere.\\
Then the gradients span the normal directions to our d-cycle. Can we 
construct in a canonical way a set of vector fields, which form a basis for 
the orthogonal complement of these normal directions?    
To answer this questions we have a look on some simple properties of 
these vector fields and very natural associated objects. We start with
a simple but important definition:
\begin{defn}{(Hamiltonian vector field)\hspace{0.05cm}}{A Hamiltonian 
              vector field $X$ is defined by the property 
              \begin{eqnarray}
                   d(X\ins\omega) =0,\nonumber
              \end{eqnarray}  
              i.e. ${\mathcal{L}}_X\omega = 0$ which reflects the property 
              of the Hamiltonian flow to preserve the symplectic form.  
              In the case of mild topology closeness implies exactness 
              and we can write   
              \begin{eqnarray}
                  df + X_f\ins\omega = 0\nonumber
              \end{eqnarray}
              assigning to the Hamiltonian vector field its generating 
              function $f$. }
\end{defn}
We will show that the Hamiltonian vector fields corresponding to the $f_i$
do span the orthogonal complement mentioned before. At first we observe 
that $X_f\perp grad\,f$ by construction. By using the ``symplectic 
involution'' $\sigma$ given by
\begin{eqnarray}
 \sigma&=&\left(\begin{array}{cc}
              0 & -\unity_d \\
              \unity_d & 0
          \end{array}\right) \;\;\;\;\;\; \sigma^2 = -\unity_{2d} \nonumber
\end{eqnarray}
we can write $X_f$ as  $X_f=\sigma\cdot grad\, f$. Sometimes $X_f$ is 
called the symplectic gradient, therefore. Now we calculate
\begin{eqnarray}
 <grad\,f,\sigma\cdot grad\,f>\,&=&\,-<\sigma^2\cdot grad\,f,
                                        \sigma\cdot grad\,f>\nonumber\\
                                 &=&\,-<\sigma\cdot grad\,f, \sigma^+\sigma
                                        \cdot grad\,f> \nonumber\\
                                 &=&\,-<grad\,f,\sigma\cdot grad\,f>\nonumber
\end{eqnarray} 
Evidently $grad\,f$ and $X_f$ are orthogonal vectors. But is $X_{f^i}$
orthogonal to all gradients $grad\,f^i$? Now we exploit the integrability 
condition. Since all functions do commute with respect to the Poisson 
bracket we conclude:
\begin{eqnarray}
0 \buildrel !\over = \{f^i,f^j\} &\buildrel Def \over 
                   =& \omega(X_{f^i},X_{f^j}) 
                   = X_{f^j}\ins X_{f^i}\ins\omega
                   = -X_{f^j}\ins df^i 
                   = -df^i(X_{f^j}) \nonumber\\
                  &=& -< grad\,f^i,\sigma\cdot grad\,f^j > \nonumber 
\end{eqnarray}
So we recognize that our integrability condition guarantees the 
orthogonality of the span $X_{f^i}$ to the normal directions. Since the 
normal span is linear independent and
\begin{eqnarray}
   <X_{f^i},X_{f^j}>\,&=&\,<\nabla\,f^i,\nabla\,f^j>\nonumber
\end{eqnarray}
the Hamiltonian vector fields are independent, too.
Now we have to care for the Lagrangian property. Does the symplectic form 
$\omega$ vanish on the subspace spanned by the $X_{f^i}$? Reading the last 
formula in the other direction 
\begin{eqnarray}
  \omega(X_{f^i},X_{f^j}) &=&  \{f^i,f^j\} \nonumber
\end{eqnarray}  
this wish becomes true, too. The next question touches the sore spot of the 
whole business. Is the space spanned by the $X_{f^i}$  tangent to the 
level surface $M_\xi$? 
We want to investigate the Hamiltonian flow generated by $f^i$. Obviously
the Hamiltonian $f^i$ is a constant of motion. Further since the other $f_j$
are in involution with $f^i$ they are constants of motion, too. Hence the 
level surface $M_\xi$ is preserved by all Hamiltonian flows corresponding 
to the associated Hamiltonian vector fields $X_{f^i}$. But for $f^i$ to be 
a constant of motion 
\begin{eqnarray}
   X_{f^j}(f^i) \buildrel ! \over = 0, \nonumber
\end{eqnarray} 
i.e. $X_{f^i}$ is tangent to $M_\xi$ everywhere. 
\begin{cor}
   The level surface $M_\xi$ is a Lagrangian submanifold.
\end{cor}

\end{appendix}

%
%

\newpage

\bibliographystyle{utphys}
\bibliography{3-cycle1}

\begingroup\raggedright\begin{thebibliography}{10}

\bibitem{witten}
E.~Witten, ``Solutions of four-dimensional field theories via {M} theory,''
  {\em Nucl. Phys.} {\bf B500} (1997) 3,
  \href{http://xxx.lanl.gov/abs/hep-th/9703166}{{\tt hep-th/9703166}}.

\bibitem{hanwit}
A.~Hanany and E.~Witten, ``Type {IIB} superstrings, {BPS} monopoles, and
  three-dimensional gauge dynamics,'' {\em Nucl. Phys.} {\bf B492} (1997)
  152--190, \href{http://xxx.lanl.gov/abs/hep-th/9611230}{{\tt
  hep-th/9611230}}.

\bibitem{sw}
N.~Seiberg and E.~Witten, ``Electric - magnetic duality, monopole condensation,
  and confinement in {N=2} supersymmetric {Yang-Mills} theory,'' {\em Nucl.
  Phys.} {\bf B426} (1994) 19--52,
  \href{http://xxx.lanl.gov/abs/hep-th/9407087}{{\tt hep-th/9407087}}.

\bibitem{hori}
K.~Hori, H.~Ooguri, and Y.~Oz, ``Strong coupling dynamics of four-dimensional
  {N=1} gauge theories from {M} theory five-brane,'' {\em Adv. Theor. Math.
  Phys.} {\bf 1} (1998) 1--52,
  \href{http://xxx.lanl.gov/abs/hep-th/9706082}{{\tt hep-th/9706082}}.

\bibitem{witten1}
E.~Witten, ``Branes and the dynamics of {QCD},'' {\em Nucl. Phys.} {\bf B507}
  (1997) 658, \href{http://xxx.lanl.gov/abs/hep-th/9706109}{{\tt
  hep-th/9706109}}.

\bibitem{9706127}
A.~Brandhuber, N.~Itzhaki, V.~Kaplunovsky, J.~Sonnenschein and 
S. Yankielowicz, ``Comments on the M theory approach to N=1 SQCD and brane
                  dynamics,''
{\em Phys. Lett.} {\bf B415} (1997) 127,
\href{http://xxx.lanl.gov/abs/hep-th/9706127}{{\tt hep-th/9706127}}.


\bibitem{hanzaf}
A.~Hanany and A.~Zaffaroni, ``On the realization of chiral four-dimensional
  gauge theories using branes,''
  \href{http://xxx.lanl.gov/abs/hep-th/9801134}{{\tt hep-th/9801134}}.

\bibitem{Landsteiner}
K.~Landsteiner,E.~Lopez and D.~A.~Lowe, ``Duality of chiral N=1 
 supersymmetric gauge theories via branes,'' 
 \href{http://xxx.lanl.gov/abs/hep-th/9801002}{{\tt hep-th/9801002}}.

\bibitem{alter}
I.~Brunner, A.~Hanany, A.~Karch, and D.~{L\"ust}, ``Brane dynamics and chiral
  nonchiral transitions,'' 
{\em Nucl. Phys.} {\bf D528} (1998) 197,
\href{http://xxx.lanl.gov/abs/hep-th/9801017}{{\tt hep-th/9801017}}.

\bibitem{gibpap}
G.~W. Gibbons and G.~Papadopoulos, ``Calibrations and intersecting branes,''
  \href{http://xxx.lanl.gov/abs/hep-th/9803163}{{\tt hep-th/9803163}}.


\bibitem{9803216},
J.~P.~Gauntlett, N.~D.~Lambert and P.~C.~West, ``Branes and 
     calibrated geometries,''
\href{http://xxx.lanl.gov/abs/hep-th/9803216}{{\tt hep-th/9803216}}.

\bibitem{kol}
B.~Kol, ``5-d field theories and {M} theory,''
  \href{http://xxx.lanl.gov/abs/hep-th/9705031}{{\tt hep-th/9705031}}.

\bibitem{gimongremm}
E.~G. Gimon and M.~Gremm, ``A note on brane boxes at finite string coupling,''
  \href{http://xxx.lanl.gov/abs/hep-th/9803033}{{\tt hep-th/9803033}}.

\bibitem{9411048}
A.~Klemm, W.~Lerche, S.~Yankielowicz and S.~Theisen,
``Simple singularities and N=2 supersymmetric Yang-Mills theory,''
{\em Phys. Lett.}, {\bf B344}, 1995, 169-175,
 \href{http://xxx.lanl.gov/abs/hep-th/9411048}{{\tt hep-th/9411048}}.

\bibitem{klemm}
A.~Klemm, W.~Lerche, S.~Yankielowicz, and S.~Theisen, ``On the monodromies of
  {N=2} supersymmetric {Yang-Mills} theory,''
  \href{http://xxx.lanl.gov/abs/hep-th/9412158}{{\tt hep-th/9412158}}.

\bibitem{argyres}
P.~C. Argyres and A.~E. Faraggi, ``The vacuum structure and spectrum of {N=2}
  supersymmetric {SU(n)} gauge theory,'' {\em Phys. Rev. Lett.} {\bf 74} (1995)
  3931--3934, \href{http://xxx.lanl.gov/abs/hep-th/9411057}{{\tt
  hep-th/9411057}}.

\bibitem{berkeley}
J.~de~Boer, K.~Hori, H.~Ooguri, and Y.~Oz, ``Kahler potential and higher
  derivative terms from {M} theory five-brane,'' {\em Nucl. Phys.} {\bf B518}
  (1998) 173, \href{http://xxx.lanl.gov/abs/hep-th/9711143}{{\tt
  hep-th/9711143}}.

\bibitem{haur}
A.~Hanany and A.~M. Uranga, ``Brane boxes and branes on singularities,''
  \href{http://xxx.lanl.gov/abs/hep-th/9805139}{{\tt hep-th/9805139}}.

\bibitem{leighstrassler}
R.~G. Leigh and M.~J. Strassler, ``Exactly marginal operators and duality in
  four-dimensional {N=1} supersymmetric gauge theory,'' {\em Nucl. Phys.} {\bf
  B447} (1995) 95--136, \href{http://xxx.lanl.gov/abs/hep-th/9503121}{{\tt
  hep-th/9503121}}.

\bibitem{HaStrUr}
A.~Hanany, M.~J. Strassler, and A.~M. Uranga, ``Finite theories and marginal
  operators on the brane,'' \href{http://xxx.lanl.gov/abs/hep-th/9803086}{{\tt
  hep-th/9803086}}.

\bibitem{aharonyhanany}
O.~Aharony and A.~Hanany, ``Branes, superpotentials and superconformal fixed
  points,'' {\em Nucl. Phys.} {\bf B504} (1997) 239,
  \href{http://xxx.lanl.gov/abs/hep-th/9704170}{{\tt hep-th/9704170}}.

\bibitem{randall}
L.~Randall, Y.~Shirman, and R.~von Unge, ``Brane boxes: Bending and beta
  functions,'' {\em Phys. Rev.} {\bf D58} (1998) 105005,
  \href{http://xxx.lanl.gov/abs/hep-th/9806092}{{\tt hep-th/9806092}}.

\bibitem{LKS}
A.~Karch, D.~{L\"ust}, and D.~Smith, ``Equivalence of geometric engineering 
  and Hanany-Witten via fractional branes,''
  \href{http://xxx.lanl.gov/abs/hep-th/9803232}{{\tt hep-th/9803232}}.

\bibitem{leighrozali}
R.~G. Leigh and M.~Rozali, ``Brane boxes, anomalies, bending and tadpoles,''
  \href{http://xxx.lanl.gov/abs/hep-th/9807082}{{\tt hep-th/9807082}}.

\bibitem{3dgauge}
O.~Aharony, A.~Hanany, K.~Intriligator, N.~Seiberg, and M.~J. Strassler,
  ``Aspects of {N=2} supersymmetric gauge theories in three- dimensions,'' {\em
  Nucl. Phys.} {\bf B499} (1997) 67,
  \href{http://xxx.lanl.gov/abs/hep-th/9703110}{{\tt hep-th/9703110}}.

\bibitem{9806177}
H.~Garcia-Compean and A.~M. Uranga, ``Brane box realization of chiral gauge
  theories in two- dimensions,''
  \href{http://xxx.lanl.gov/abs/hep-th/9806177}{{\tt hep-th/9806177}}.

\bibitem{BBS}
K.~Becker, M.~Becker and A.~Strominger, ``Five-branes, membranes 
and nonperturbative string theory,'' \href{http://xxx.lanl.gov/abs/hep-th/9507158}{{\tt 9507158}}.

\bibitem{HL}
R.~Harvey and H.~B.~Lawson ``Calibrated geometries,'' Acta Mathematica 148.

\bibitem{AB}
A.~Armoni and A.~Brandhuber, ``Comments on (non)chiral gauge theories and type
  IIB branes,'' \href{http://xxx.lanl.gov/abs/hep-th/9803186}{{\tt
  hep-th/9803186}}.

\bibitem{workinpro}
A.~Karch, D.~{L\"ust} and A.~Miemiec, work in progress.

\end{thebibliography}\endgroup



%


\end{document}